\documentclass[12pt,a4paper]{article}
\pdfoutput=1
\usepackage{amsmath,amssymb,bbm,slashed,cancel}
\usepackage{graphicx}
\usepackage{slashed}
\usepackage{hyperref}
\usepackage{subcaption}
\usepackage{cite}
\usepackage{dsfont}
\usepackage{stackengine}
\usepackage{accents}
\usepackage[utf8]{inputenc}
\usepackage{multirow}

\usepackage{colortbl,colordvi,xcolor,color}

\usepackage{array,multirow}
\usepackage{booktabs}
\usepackage{float}

\allowdisplaybreaks

\newcommand{\eg}{{\em e.g.}}
\newcommand{\ie}{{\em i.e.}}

\newcommand\unit[1]{\,\mathrm{#1}}
\newcommand\MeV{\unit{MeV}}
\newcommand\GeV{\unit{GeV}}
\newcommand\TeV{\unit{TeV}}
\newcommand\olrarrow[1]{\overset{\text{\scriptsize$\leftrightarrow$}}{#1}}

\newcommand{\crowcolor}{\rowcolor[rgb]{0.9,0.9,0.9}}

\makeatletter
\newcommand\xleftrightarrow[2][]{%
  \ext@arrow 9999{\longleftrightarrowfill@}{#1}{#2}}
\newcommand\longleftrightarrowfill@{%
  \arrowfill@\leftarrow\relbar\rightarrow}
\makeatother

\begin{document}

\begin{titlepage}
\setcounter{page}{0} 
\begin{flushright}
KEK--TH--2412
\end{flushright}
\vskip 1.5cm
\begin{center}
  {\Large \bf New physics interpretation of $W$-boson mass anomaly}
\vskip 1.5cm
{
  Motoi Endo$^{(a,b)}$
  and
  Satoshi Mishima$^{(a)}$
}

\vspace{1.5em}

\begingroup\small\itshape
$^{(a)}$ \!\! KEK Theory Center, IPNS, KEK, Tsukuba, Ibaraki 305--0801, Japan
\\[0.3em]
$^{(b)}$ \!\! The Graduate University of Advanced Studies (Sokendai), Tsukuba, Ibaraki 305--0801, Japan
\endgroup

\vspace{1cm}

\abstract{
\noindent 
The CDF collaboration has recently reported an updated result on the $W$-boson mass measurement, showing a $7\sigma$ deviation from the standard model prediction.
The discrepancy may indicate new contributions to the Fermi coupling constant. 
We study simple extensions of the standard model by introducing an extra scalar, fermion or vector field.
It is found that the tension implies the new physics existing in multi-TeV scales if the new coupling to the electron and/or muon is of order unity. 
}

\end{center}
\end{titlepage}

\renewcommand{\thefootnote}{\#\arabic{footnote}}
\setcounter{footnote}{0}

\section{Introduction}

The CDF collaboration has very recently reported an updated result of the $W$-boson mass~\cite{CDF:2022hxs},
\begin{align}
M_W = (80.4335 \pm 0.0064_{\rm stat} \pm 0.0069_{\rm syst}) \GeV = 80.4335 \pm 0.0094 \GeV.
\end{align}
The result shows a deviation at more than $7\sigma$ level from the Standard Model (SM) prediction, $M_W =(80.3500 \pm 0.0056)\GeV$~\cite{deBlas:2022hdk}.
By combining the experimental results of $M_W$ from LEP 2, Tevatron~\cite{CDF:2022hxs}, LHC ATLAS~\cite{ATLAS:2017rzl} and LHCb~\cite{LHCb:2021bjt}, the averaged value is~(see, \eg, Ref.~\cite{deBlas:2022hdk})
\begin{align}
M_W = 80.4133 \pm 0.0080 \GeV,
\end{align}
and the deviation becomes $6.5\sigma$.
Although the CDF result has a tension with the previous experimental data as well as the SM prediction, if this discrepancy would be confirmed in future, it might be a sign of new physics beyond the SM. Implications of the discrepancy have been studied in Refs.~\cite{Fan:2022dck,Zhu:2022tpr,Lu:2022bgw,Athron:2022qpo,Yuan:2022cpw,Strumia:2022qkt,Yang:2022gvz,deBlas:2022hdk,Du:2022pbp,Tang:2022pxh,Cacciapaglia:2022xih,Blennow:2022yfm,Arias-Aragon:2022ats,Zhu:2022scj,Sakurai:2022hwh,Fan:2022yly,Liu:2022jdq,Lee:2022nqz,Cheng:2022jyi,Song:2022xts,Bagnaschi:2022whn,Paul:2022dds,Bahl:2022xzi,Asadi:2022xiy,DiLuzio:2022xns,Athron:2022isz,Gu:2022htv,Heckman:2022the,Babu:2022pdn}. 

The SM prediction of the $W$-boson mass is determined by the electroweak precision observables (EWPO). 
The CDF discrepancy implies the following three possibilities in terms of the SM effective field theory (SMEFT~\cite{Grzadkowski:2010es}); i) new physics contributions arising in the operator $(\phi^\dagger\sigma^a\phi W^a_{\mu\nu}B^{\mu\nu})$, ii) those appearing in $(\phi^\dagger D_\mu \phi)((D^\mu\phi)^\dagger\phi)$ and iii) those via the Fermi coupling constant. 
Here, $\phi$ is the SM Higgs boson, $W^a_{\mu\nu}$ ($B_{\mu\nu}$) is the field strength of the $\mathrm{SU(2)}_L$ ($\mathrm{U(1)}_Y$) gauge boson, and $D_\mu$ is the covariant derivative.
In other words, the contributions to the first and second operators are understood as new physics effects on the oblique $S$ and $T$ parameters, respectively. See, {\it e.g.},  Refs.~\cite{Fan:2022dck,Lu:2022bgw,Athron:2022qpo,Yuan:2022cpw,Strumia:2022qkt,deBlas:2022hdk,Du:2022pbp,Cacciapaglia:2022xih,Sakurai:2022hwh,Fan:2022yly,Liu:2022jdq,Song:2022xts,Bagnaschi:2022whn,Paul:2022dds,Asadi:2022xiy,DiLuzio:2022xns,Gu:2022htv,Babu:2022pdn} for such studies in light of the CDF result.
Alternatively, we study the third possibility in this paper.\footnote{In Ref.~\cite{Blennow:2022yfm}, the $W$-boson mass discrepancy is resolved by a shift in the Fermi constant caused by right-handed neutrinos.}
The Fermi coupling constant $G_F$ is determined precisely by measuring the muon decay to the electron, and receives new physics corrections of $(\phi^\dagger i \olrarrow{D_\mu^a} \phi) (\bar\ell_i \gamma^\mu \sigma^a \ell_j)$ and $(\bar\ell_i \gamma_\mu \ell_j) (\bar\ell_k \gamma^\mu \ell_l)$, where $\ell_i$ is a left-handed lepton in the $i$-th generation. 

In this paper, we discuss new physics scenarios that affect $G_F$.
In particular, we consider simple extensions of the SM by introducing an extra scalar, fermion or vector field.
Among them, scalar fields with a hypercharge $1$ and with couplings to the SM leptons can contribute to $G_F$ via the four-Fermi interactions $(\bar\ell_i \gamma_\mu \ell_j) (\bar\ell_k \gamma^\mu \ell_l)$.
Also, extra vector bosons which have charged-current interactions with the SM leptons may mimic the SM $W$ boson and contribute to $G_F$.
On the other hand, extra leptons which couple to the SM Higgs boson as well as the SM leptons induce SMEFT operators including $(\phi^\dagger i \olrarrow{D_\mu^a} \phi) (\bar\ell_i \gamma^\mu \sigma^a \ell_j)$.
Thus, these fields may change the $W$-boson mass via $G_F$.
In this paper, we also examine flavor-dependent contributions of new physics. 

This paper is organized as follows.
In Sections~\ref{sec:EWPO} we explain constraints from the EWPO and study the updated result on the $W$-boson mass measurement in the SMEFT framework.
In Section~\ref{sec:model} we investigate single-field extensions of the SM.
Finally our conclusions are drawn in Section~\ref{sec:conclusion}.

\section{Electroweak precision observables}
\label{sec:EWPO}

The EWPO including the $W$-boson mass $M_W$ receive contributions of new physics.
If its energy scale is higher than the electroweak scale, they are represented in terms of higher dimensional operators of the SMEFT,
\begin{align}
 \mathcal{L}_{d>4} &= \sum_i C_i \mathcal{O}_i,
\end{align}
where $C_i$ is a Wilson coefficient and $\mathcal{O}_i$ is a higher dimensional operator. 
The dimension-six operators relevant for $M_W$ in our scenarios are 
\begin{align}
 (\mathcal{O}_{\phi\ell}^{(3)})_{ij} &= 
 (\phi^\dagger i \olrarrow{D_\mu^a} \phi) (\bar\ell_i \gamma^\mu \sigma^a \ell_j), \\
 (\mathcal{O}_{\ell\ell})_{ijkl} &= 
 (\bar\ell_i \gamma_\mu \ell_j) (\bar\ell_k \gamma^\mu \ell_l),
\end{align}
where $\ell_i$ $(e_{Ri})$ denotes the $\mathrm{SU(2)}_L$ doublet (singlet) lepton in the $i$-th generation, and the derivatives mean
\begin{align}
 \phi^\dagger \olrarrow{D_\mu} \phi =
 \phi^\dagger (D_\mu \phi) - (D_\mu \phi)^\dagger \phi,~~~~~
 \phi^\dagger \olrarrow{D_\mu^a} \phi =
 \phi^\dagger \sigma^a (D_\mu \phi) - (D_\mu \phi)^\dagger \sigma^a \phi,
\end{align}
with the Pauli matrix $\sigma^a$.
In addition, when fermion extensions of the SM are discussed in the next section, the EWPO and the leptonic decay of the Higgs boson are affected via the operators,
\begin{align}
 (\mathcal{O}_{\phi\ell}^{(1)})_{ij} &= 
 (\phi^\dagger i \olrarrow{D_\mu} \phi) (\bar\ell_i \gamma^\mu \ell_j), \\
 (\mathcal{O}_{e\phi})_{ij} &= (\phi^\dagger\phi)
 (\bar\ell_i\hspace{0.2mm}  \phi\hspace{0.5mm} e_{Rj}).
\end{align}
Let us define dimensionless coefficients as 
\begin{align}
 \widehat C_i = v^2\, C_i. 
\end{align}
Here $v$ denotes the vacuum expectation value (VEV) of the Higgs field, $\phi = \left[0,(v+h)/\sqrt{2}\right]^T$, where the Nambu–Goldstone bosons are ignored.

The operator $\mathcal{O}_{\phi\ell}^{(3)}$ alters the charged-current interactions of leptons after the electroweak symmetry breaking, and the four-Fermi operator $\mathcal{O}_{\ell\ell}$ contributes directly to the muon decay to the electron and neutrinos. 
Therefore, the measured value of $G_F$ from the decay is shifted from the SM prediction as 
\begin{align}
 G_F =
 \frac{1}{\sqrt{2}\,v^2}
 \left( 1 + \delta_{G_F} \right),~~~
 \delta_{G_F} = 
 (\widehat{C}_{\phi\ell}^{(3)})_{11} + (\widehat{C}_{\phi\ell}^{(3)})_{22} - (\widehat{C}_{\ell\ell})_{1221}.
 \label{eq:GF}
\end{align}
Note that $(\widehat{C}_{\ell\ell})_{1221} = (\widehat{C}_{\ell\ell})_{2112}$.
Then, the modification of $G_F$ affects the $W$-boson mass as\footnote{
Here, we omitted contributions from $\mathcal{O}_{\phi WB} = (\phi^\dagger\sigma^a\phi W^a_{\mu\nu}B^{\mu\nu})$ and $\mathcal{O}_{\phi D} = (\phi^\dagger D_\mu \phi)((D^\mu\phi)^\dagger\phi)$, which can be taken into account by replacing
$\delta_{G_F} \to \delta_{G_F} + 2\frac{c_W}{s_W} \widehat{C}_{\phi WB} + \frac{c_W^2}{2s_W^2} \widehat{C}_{\phi D}$ in Eq.~\eqref{eq:mW}.
}
\begin{align}
 M_W = (M_W)_{\rm SM} 
 \left[ 1 - \frac{s_W^2}{2(c_W^2-s_W^2)} \delta_{G_F} \right], 
 \label{eq:mW}
\end{align}
where $s_W$ and $c_W$ are the sine and cosine of the weak mixing angle. 
A quantity with the subscript ``SM'' denotes the SM prediction, which is calculated with the measured values of the input parameters $G_F$, $\alpha$, $M_Z$, {\em etc}. 
The $W$-boson partial widths also receive the corrections to $M_W$ and those to the charged-current couplings as
\begin{align}
 \Gamma(W^+ \to \ell_i^+ \nu_{\ell i})
 &= 
 \Gamma(W^+ \to \ell_i^+ \nu_{\ell i})_{\mathrm{SM}}
 \left[
 1 - \frac{1+c_W^2}{2(c_W^2-s_W^2)}\, \delta_{G_F} 
 + 2\,(\widehat{C}_{\phi\ell}^{(3)})_{ii}
 \right], 
 \label{eq:Wwidth} \\
 \Gamma(W^+ \to ij)
 &=
 \Gamma(W^+ \to ij)_{\mathrm{SM}}
 \left[
 1 - \frac{1+c_W^2}{2(c_W^2-s_W^2)}\, \delta_{G_F}
 \right].
\end{align}
where $ij$ in the second equation represents quark final states such as $\bar{d}u$ and $\bar{s}c$.

The operators $\mathcal{O}_{\phi\ell}^{(1)}$ and $\mathcal{O}_{\phi\ell}^{(3)}$ contribute to the neutral-current interactions of left-handed leptons. 
The $Z$-boson couplings to the SM fermions $f$ are written as
\begin{align}
 \mathcal{L}_Z
 &=
 \frac{g}{c_W}\, 
 \bar{f} \gamma^\mu 
 \Big[
 (T_L^{\prime 3} - Q s_W^2 + \delta g_L) P_L
 + 
 (T_R^{\prime 3} - Q s_W^2 + \delta g_R) P_R  \Big] f\,Z_\mu, 
\end{align}
where $T_{L,R}^{\prime 3}$ and $Q$ are the weak isospin and the electric charge of $f$.
The new physics contributions are obtained as
\begin{align}
 \delta g_L &=
 \begin{cases}
 \displaystyle
 - \frac{1}{2}
   \left[ T_L^{\prime 3} + \frac{Q s_W^2}{c_W^2-s_W^2} \right] \delta_{G_F}
 - \frac{1}{2}\, (\widehat{C}_{\phi \ell}^{(1)})_{ii}
 + T_L^{\prime 3}\, (\widehat{C}_{\phi \ell}^{(3)})_{ii}
 & \mathrm{for}~~f=\ell_i, \nu_{\ell i},
 \label{eq:Zcoupling} \\[10pt]
\displaystyle
 - \frac{1}{2}
   \left[ T_L^{\prime 3} + \frac{Q s_W^2}{c_W^2-s_W^2} \right] \delta_{G_F}
 & \mathrm{otherwise},
 \end{cases}
 \\[5pt]
 \delta g_R &=
 - \frac{Q s_W^2}{2(c_W^2-s_W^2)}\, \delta_{G_F}.
\end{align}
The $Z$-boson observables in Table~\ref{tab:EWPO} are represented in terms of these effective $Zff$ couplings (see, \eg,\ Ref.~\cite{Ciuchini:2013pca}). 

We perform a Bayesian fit of the SMEFT operators to the experimental data of the EWPO~\cite{ALEPH:2005ab,Schael:2013ita,Janot:2019oyi}.
The analysis utilizes the \texttt{HEPfit v1.0} package~\cite{deBlas:2019okz}, which is based on the Markov Chain Monte Carlo provided by the Bayesian Analysis Toolkit (\texttt{BAT})~\cite{Caldwell:2008fw}. 
The full two-loop electroweak corrections are included for the SM contributions to $M_W$ and the $Z$-boson observables~\cite{Awramik:2003rn,Awramik:2006uz,Dubovyk:2019szj}, while the $W$-boson widths are calculated at one-loop level~\cite{Bardin:1986fi,Denner:1990tx}. 
Additionally new physics contributions are implemented to the package for the current work. 
Theoretical uncertainties from missing higher-order corrections in the SM are included only for the $W$ mass as $\delta_{\mathrm{th}} M_W = 0\pm 4\MeV$~\cite{Awramik:2003rn} assuming the Gaussian distribution, while those for the other observables are not significant~\cite{Dubovyk:2018rlg} and are neglected from the fit. 
The input values necessary for this study are summarized in Table~\ref{tab:EWPO}, where the the choice of $\alpha_s(M_Z^2)$, $\Delta\alpha_{\mathrm{had}}^{(5)}(M_Z^2)$ and $m_t$ is followed by those in Refs.~\cite{deBlas:2021wap,deBlas:2022hdk}. 
In the following analysis, the parameters $G_F$, $\alpha$ and the light fermion masses are fixed to be constants.
\begin{table}[t]
\centering
\begin{tabular}{ccc|ccc}
\hline
& Measurement & Ref. &
& Measurement & Ref.
\\
\hline
$\alpha_s(M_Z^2)$ & 
$0.1177 \pm 0.0010$ &
\cite{deBlas:2021wap} &
$M_Z$ [GeV] &
$91.1876 \pm 0.0021$ &
\cite{Janot:2019oyi} 
\\
\cline{1-3}
$\Delta\alpha_{\mathrm{had}}^{(5)}(M_Z^2)$ &
$0.02766 \pm 0.00010$ &
\cite{deBlas:2021wap} &
$\Gamma_Z$ [GeV] &
$2.4955 \pm 0.0023$ &
\\
\cline{1-3}
$m_t$ [GeV] &
$171.79 \pm 0.38$ &
\cite{deBlas:2022hdk} &
$\sigma_{h}^{0}$ [nb] &
$41.4807 \pm 0.0325$ &
\\
\cline{1-3}
$m_h$ [GeV] &
$125.21 \pm 0.12$ &
\cite{deBlas:2021wap} &
$R^{0}_{e}$ &
$20.8038 \pm 0.0497$ &
\\
\cline{1-3}
$M_W$ [GeV] &
$80.4133 \pm 0.0080$ &
&
$R^{0}_{\mu}$ &
$20.7842 \pm 0.0335$ &
\\
\cline{1-3}
$\Gamma_{W}$ [GeV] & 
$2.085 \pm 0.042$ &
\cite{ParticleDataGroup:2020ssz} &
$R^{0}_{\tau}$ &
$20.7644 \pm 0.0448$ &
\\
\cline{1-3}
$\mathcal{B}(W\to e\nu)$ &
$0.1071 \pm 0.0016$ &
\cite{Schael:2013ita} &
$A_{\mathrm{FB}}^{0, e}$ &
$0.0145 \pm 0.0025$ &
\\
$\mathcal{B}(W\to \mu\nu)$ &
$0.1063 \pm 0.0015$ &
&
$A_{\mathrm{FB}}^{0, \mu}$ &
$0.0169 \pm 0.0013$ &
\\
$\mathcal{B}(W\to \tau\nu)$ &
$0.1138 \pm 0.002$ &
&
$A_{\mathrm{FB}}^{0, \tau}$ &
$0.0188 \pm 0.0017$ &
\\
\hline
$R(\tau/\mu)$ &
$0.992 \pm 0.013$ &
\cite{ATLAS:2020xea} &
$R^{0}_{b}$ &  
$0.21629 \pm 0.00066$ & 
\cite{ALEPH:2005ab,Bernreuther:2016ccf}
\\
\cline{1-3}
$\mathcal{A}_e$ (SLD) & 
$ 0.1516 \pm 0.0021 $ &
\cite{ALEPH:2005ab} &
$R^{0}_{c}$ & 
$0.1721 \pm 0.0030$ & 
\\
$\mathcal{A}_\mu$ (SLD) & 
$ 0.142 \pm 0.015 $ &
&
$A_{\mathrm{FB}}^{0, b}$ & 
$0.0996 \pm 0.0016$ &
\\
$\mathcal{A}_\tau$ (SLD) & 
$ 0.136 \pm 0.015 $ &
&
$A_{\mathrm{FB}}^{0, c}$ & 
$0.0707 \pm 0.0035$ &
\\
\cline{1-3}
$\mathcal{A}_e$ (LEP) & 
$ 0.1498 \pm 0.0049 $ &
\cite{ALEPH:2005ab}
&
$\mathcal{A}_b$ & 
$0.923 \pm 0.020$ &
\\
$\mathcal{A}_\tau$ (LEP) & 
$ 0.1439 \pm 0.0043 $ &
&
$\mathcal{A}_c$ & 
$0.670 \pm 0.027$ &
\\
\hline
\end{tabular}
\caption{Experimental measurement of the SM input parameters and EWPO.}
\label{tab:EWPO}
\end{table}

\begin{figure}[t]
\centering
\includegraphics[scale=0.7]{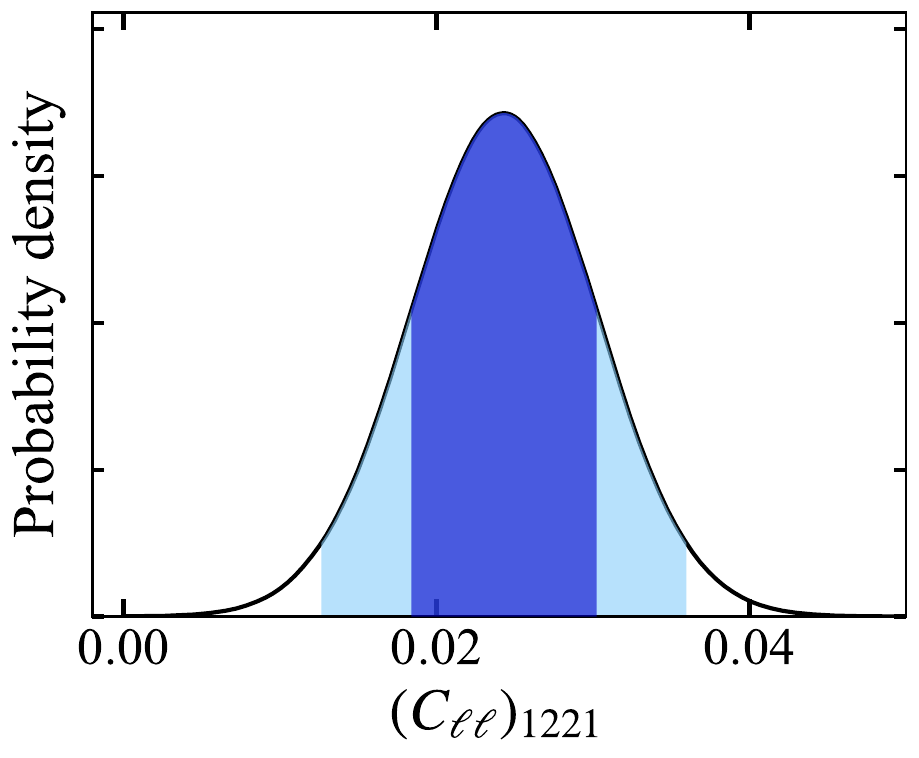}
\caption{Probability distribution for $(C_{\ell\ell})_{1221}$ in units of $\TeV^{-2}$ obtained from a fit to the EWPO, where the darker (lighter) region corresponds to the 68\% (95\%) probability.} 
\label{fig:CLL_1221}
\end{figure}

\begin{figure}[t]
\centering
\includegraphics[scale=0.7]{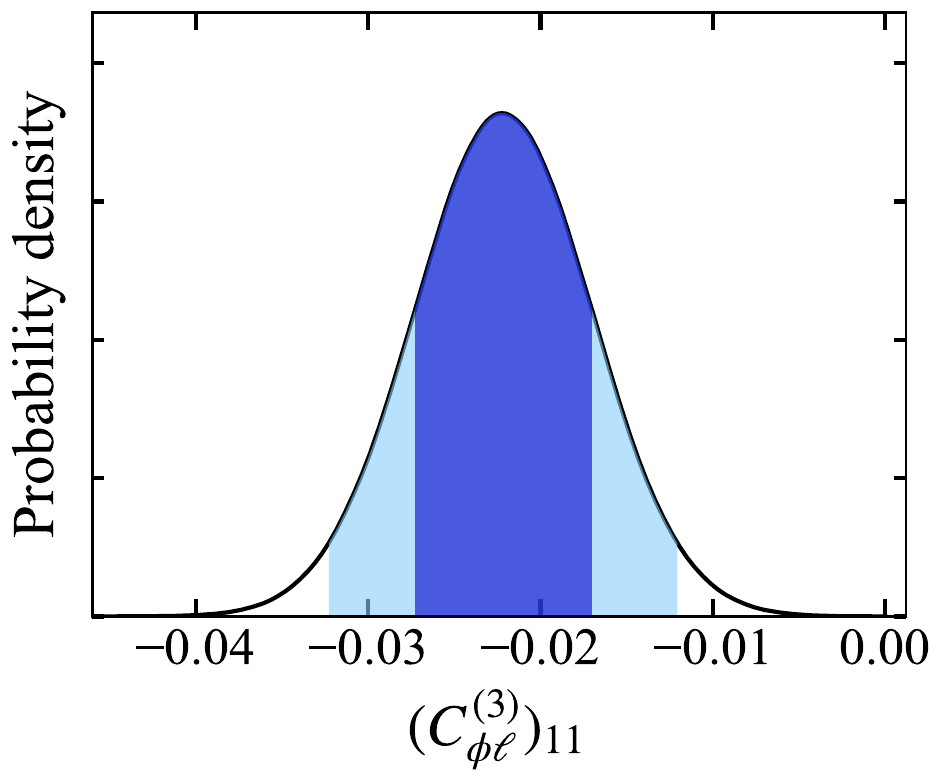}
\hspace{10mm}
\includegraphics[scale=0.7]{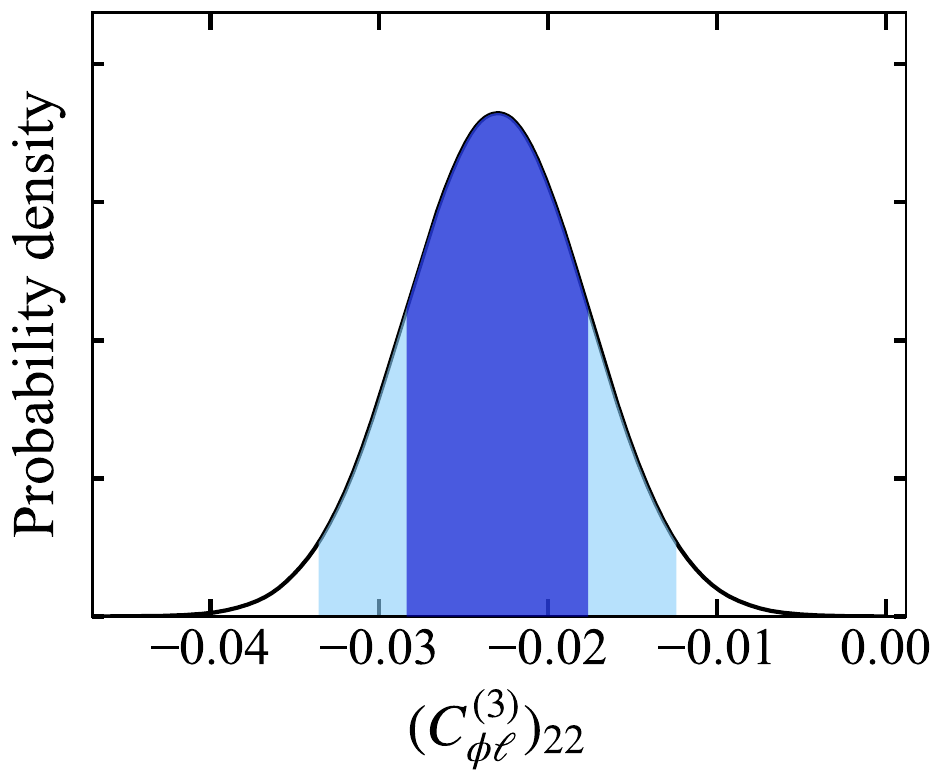}
\\[1mm]
\includegraphics[scale=0.7]{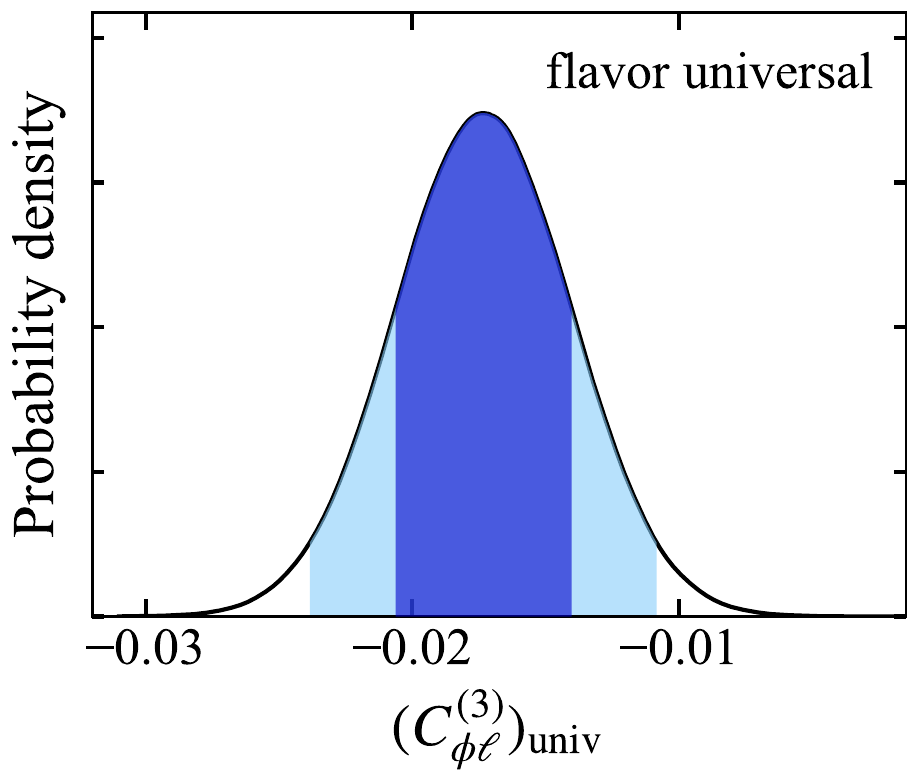}
\hspace{3mm}
\includegraphics[scale=0.7]{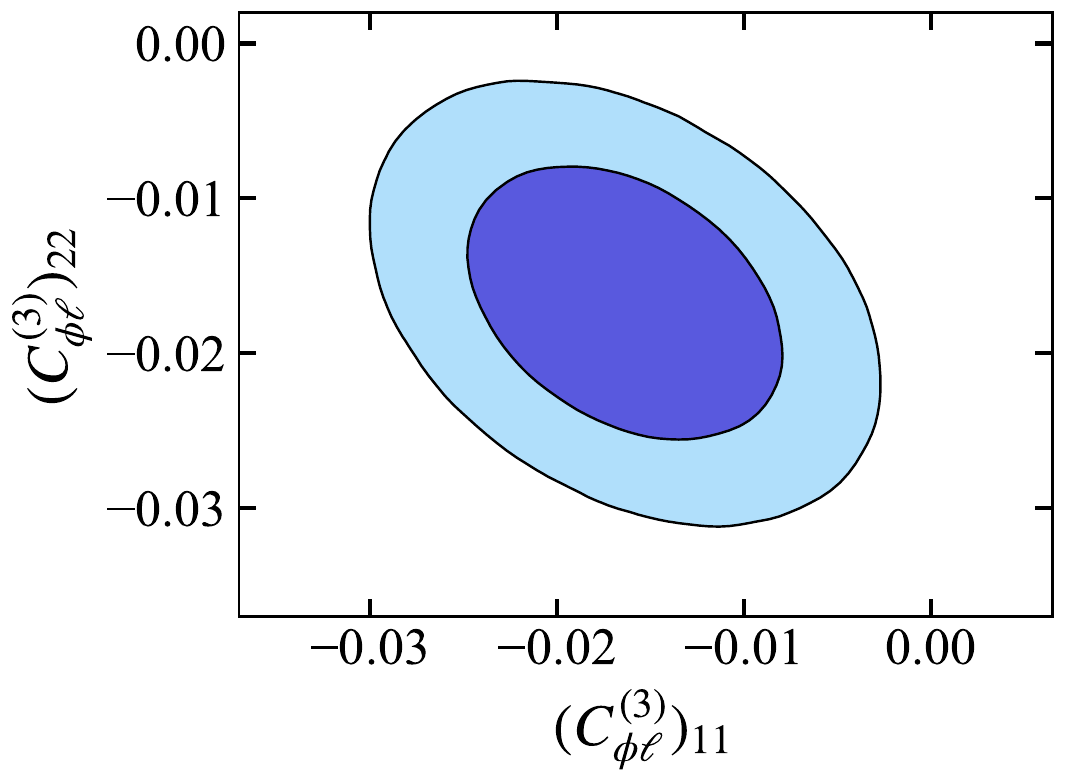}
\caption{Probability distributions for the coefficients $C_{\phi\ell}^{(3)}$. The horizon axis is shown in units of $\TeV^{-2}$. Only a single operator is switched on at a time for the top two plots, while the lepton-flavor universality is assumed for the bottom-left plot: $(C_{\phi \ell}^{(3)})_{\mathrm{univ}}\equiv 
(C_{\phi \ell}^{(3)})_{11} = (C_{\phi \ell}^{(3)})_{22} = (C_{\phi \ell}^{(3)})_{33}$. 
In the bottom-right plot, the two coefficients are fitted simultaneously.}
\label{fig:CHL3}
\end{figure}

As shown in Eq.~\eqref{eq:GF}, the Fermi coupling constant is affected by $(C_{\phi\ell}^{(3)})_{11,22}$ and $(C_{\ell\ell})_{1221}$.
In Figs.~\ref{fig:CLL_1221} and \ref{fig:CHL3}, we show the probability distributions for $(C_{\ell\ell})_{1221}$ and $(C_{\phi\ell}^{(3)})_{ii}$, respectively.
The horizontal axes are shown in units of $\TeV^{-2}$.
It is noticed from Eq.~\eqref{eq:GF} that $G_F$ depends on $C_{\phi\ell}^{(3)}$ in a combination of $(C_{\phi\ell}^{(3)})_{11} + (C_{\phi\ell}^{(3)})_{22}$. 
According to Eqs.~\eqref{eq:Wwidth} and \eqref{eq:Zcoupling}, these Wilson coefficients also affect the $Z$-observables and leptonic decays of the $W$-boson directly.
Hence we performed the analysis in a flavor-dependent way. 
For the top two plots in Fig.~\ref{fig:CHL3}, only a single operator for $(C_{\phi\ell}^{(3)})_{11}$ or $(C_{\phi\ell}^{(3)})_{22}$ is switched on, while the lepton-flavor universality is assumed for the bottom-left plot: $(C_{\phi \ell}^{(3)})_{\mathrm{univ}}\equiv 
(C_{\phi \ell}^{(3)})_{11} = (C_{\phi \ell}^{(3)})_{22} = (C_{\phi \ell}^{(3)})_{33}$. 
In the bottom-right plot, the two coefficients $(C_{\phi\ell}^{(3)})_{11}$ and $(C_{\phi\ell}^{(3)})_{22}$ are fitted simultaneously.
From the figures it is found that the Wilson coefficient is implied to take a positive value for $(C_{\ell\ell})_{1221}$, while $(C_{\phi \ell}^{(3)})_{\mathrm{univ}} < 0$ is favored to relax the $W$-boson mass discrepancy. 
The numerical values of the 68\% and 95\% probability ranges are summarized in Tables \ref{tab:num} and \ref{tab:num2}. 
As will be discussed in the next section, these results are useful to discriminate the models.

\begin{table}[t]
\centering
\begin{tabular}{ccc}
\hline
Coefficient & 68\% prob. range & 95\% prob. range 
\\
\hline
$(C_{\ell\ell})_{1221}$ &
$[0.018,\ 0.030]$ &
$[0.013,\ 0.036]$
\\
$(C_{\phi \ell}^{(3)})_{11}$ &
$[-0.027,\ -0.017]$ &
$[-0.032,\ -0.012]$ 
\\
$(C_{\phi \ell}^{(3)})_{22}$ &
$[-0.028,\ -0.018]$ &
$[-0.034,\ -0.012]$ 
\\
$(C_{\phi \ell}^{(3)})_{\mathrm{univ}}$ &
$[-0.021,\ -0.014]$ &
$[-0.024,\ -0.011]$ 
\\
\hline
$|(y_{\Xi_1})_{12}|$ &
$[0.14,\ 0.17]$ &
$[0.11,\ 0.19]$ 
\\
$\mathcal{G}_{\mathcal{W}}$ &
$[0.27,\ 0.35]$ &
$[0.22,\ 0.38]$ 
\\
$|(g_{\mathcal{W}})_{12}|$ &
$[0.38,\ 0.49]$ &
$[0.32,\ 0.54]$ 
\\
\hline
$|(\lambda_{E})_{1}|$ &
$[0.13,\ 0.20]$ &
$[0.08,\ 0.23]$ 
\\
$|(\lambda_{E})_{2}|$ &
$[0.23,\ 0.30]$ &
$[0.19,\ 0.33]$ 
\\
$|(\lambda_{E})_{\mathrm{univ}}|$ &
$[0.17,\ 0.22]$ &
$[0.14,\ 0.24]$ 
\\
\hline
\end{tabular}
\caption{Fit results for the SMEFT coefficients 
in units of TeV$^{-2}$ 
and for the new couplings 
at 68\% and 95\% probability ranges. 
Here $\mathcal{G}_{\mathcal{W}} = \sqrt{-(g_{\mathcal{W}})_{11}(g_{\mathcal{W}})_{22}}$ and $M_{\Xi_1, \mathcal{W}, E} = 1\TeV$.
In the SMEFT fits, only a single operator is switched on at a time, while 
the lepton-flavor universality is assumed 
for $(C_{\phi \ell}^{(3)})_{\mathrm{univ}}$ and $(\lambda_{E})_{\mathrm{univ}}$, {\it i.e.}, 
$(C_{\phi \ell}^{(3)})_{\mathrm{univ}}\equiv 
(C_{\phi \ell}^{(3)})_{11} = (C_{\phi \ell}^{(3)})_{22} = (C_{\phi \ell}^{(3)})_{33}$
and 
$(\lambda_{E})_{\mathrm{univ}} \equiv (\lambda_{E})_{1} = (\lambda_{E})_{2} = (\lambda_{E})_{3}$. }
\label{tab:num}
\end{table}
\begin{table}[t]
\centering
\begin{tabular}{ccc}
\hline
Coefficient & 68\% prob. range & 95\% prob. range 
\\
\hline
$(C_{\phi \ell}^{(3)})_{11}$ &
$[-0.022,\ -0.011]$ &
$[-0.027,\ -0.005]$ 
\\
$(C_{\phi \ell}^{(3)})_{22}$ &
$[-0.023,\ -0.011]$ &
$[-0.028,\ -0.005]$ 
\\
\hline
$|(\lambda_{E})_{1}|$ &
$[0.13,\ 0.20]$ &
$[0.07,\ 0.23]$ 
\\
$|(\lambda_{E})_{2}|$ &
$[0.23,\ 0.29]$ &
$[0.18,\ 0.32]$ 
\\
\hline
\end{tabular}
\caption{Same as Table~\ref{tab:num}, but for the fits with the two coefficients $(C_{\phi \ell}^{(3)})_{11}$ and $(C_{\phi \ell}^{(3)})_{22}$, or $|(\lambda_{E})_{1}|$ and $|(\lambda_{E})_{2}|$.}
\label{tab:num2}
\end{table}

Table \ref{tab:pull} shows pull values for the individual parameters and observables in each fit together with the values of the information criterion (IC) defined by 
\begin{align}
IC
&=
-2\hspace{0.5mm}\overline{\,\ln L}
+ 4\hspace{0.5mm}\sigma^2_{\ln L},
\end{align}
where $\overline{\,\ln L}$ and $\sigma^2_{\ln L}$ are 
the posterior mean and the variance of the log-likelihood distribution, respectively~\cite{Ando:2011}.
The IC is a measure of relative goodness of fit; preferred scenarios give smaller $IC$ values. 
It is found that all the SMEFT scenarios considered in Figs.~\ref{fig:CLL_1221} and \ref{fig:CHL3} give smaller $IC$ values compared to the SM, and therefore we conclude that they are preferred over the SM. 
The quality of each SMEFT fit is similar to each other. 
In those scenarios the $W$-boson mass discrepancy is relaxed by the new physics contributions to $(C_{\phi\ell}^{(3)})_{11,22}$ or $(C_{\ell\ell})_{1221}$, but the pull values for $M_W$ cannot be smaller than the two sigma level due to the increases of the pull for other observables. 
Especially, 
in all the scenarios, the larger pulls for $M_W$ and $\mathcal{A}_e$~(SLD) observed in the SM fit are reduced by the new physics contributions, but that for $A_{\mathrm{FB}}^{0, b}$ is worsened at the same time. 
\begin{table}[t]
\centering
\begin{tabular}{c|r|r|rrrr|rrrr}
\hline
&
\multirow{2}{*}{SM} &
$C_{\ell\ell}$\hspace{1mm} &
\multicolumn{4}{c|}{$C_{\phi \ell}^{(3)}$} &
\multicolumn{4}{c}{$|\lambda_{E}|$} 
\\
&&
$1221$ &
$11$ &
$22$ &
univ &
$11,22$ &
$1$ &
$2$ &
univ &
$1,2$ 
\\
\hline
$IC$ & 
86 & 65 & 65 & 64 & 54 & 58 & 81 & 67 & 67 & 65
\\
\hline
$\alpha_s(M_Z^2)$
& $-0.1$& $0.1$& $0.5$& $0.2$& $0.5$& $0.5$& $0.3$& $0.3$& $0.6$& $0.6$
\\
$\Delta\alpha_{\mathrm{had}}^{(5)}(M_Z^2)$
& $0.9$& $0.2$& $0.4$& $0.2$& $0.0$& $0.1$& $0.8$& $0.4$& $0.5$& $0.3$
\\
$m_t$
& $-1.1$& $-0.5$& $-0.7$& $-0.6$& $-0.4$& $-0.4$& $-1.0$& $-0.7$& $-0.8$& $-0.6$
\\
$m_h$
& $0.0$& $0.0$& $0.0$& $0.0$& $0.0$& $0.0$& $0.0$& $0.0$& $0.0$& $0.0$
\\
$M_W$
& $4.6$& $2.9$& $2.9$& $3.0$& $2.1$& $2.2$& $4.0$& $3.4$& $3.1$& $2.9$
\\
$\delta_{\mathrm{th}} M_W$
& $-2.0$& $-1.3$& $-1.3$& $-1.3$& $-1.0$& $-1.0$& $-1.7$& $-1.5$& $-1.4$& $-1.3$
\\
$\Gamma_{W}$
& $-0.1$& $-0.2$& $-0.2$& $-0.2$& $-0.2$& $-0.2$& $-0.1$& $-0.2$& $-0.1$& $-0.2$
\\
$\mathcal{B}(W\to e\nu)$
& $-0.8$& $-0.8$& $-0.7$& $-0.8$& $-0.7$& $-0.7$& $-0.8$& $-0.8$& $-0.8$& $-0.8$
\\
$\mathcal{B}(W\to \mu\nu)$
& $-1.4$& $-1.4$& $-1.4$& $-1.2$& $-1.3$& $-1.3$& $-1.4$& $-1.3$& $-1.3$& $-1.3$
\\
$\mathcal{B}(W\to \tau\nu)$
& $2.6$& $2.6$& $2.6$& $2.6$& $2.6$& $2.5$& $2.6$& $2.6$& $2.6$& $2.6$
\\
$R(\tau/\mu)$ 
& $-0.6$& $-0.6$& $-0.6$& $-0.8$& $-0.6$& $-0.8$& $-0.6$& $-0.8$& $-0.6$& $-0.8$
\\
$\mathcal{A}_e$ (SLD)
& $2.0$& $0.4$& $1.7$& $0.5$& $0.6$& $0.7$& $2.2$& $0.9$& $1.7$& $1.1$
\\
$\mathcal{A}_\mu$ (SLD)
& $-0.4$& $-0.6$& $-0.6$& $-0.4$& $-0.5$& $-0.5$& $-0.4$& $-0.3$& $-0.4$& $-0.3$
\\
$\mathcal{A}_\tau$ (SLD)
& $-0.8$& $-1.0$& $-1.0$& $-1.0$& $-0.9$& $-1.1$& $-0.8$& $-0.9$& $-0.8$& $-1.0$
\\
$\mathcal{A}_e$ (LEP)
& $0.5$& $-0.2$& $0.4$& $-0.1$& $-0.1$& $-0.1$& $0.6$& $0.0$& $0.4$& $0.1$
\\
$\mathcal{A}_\tau$ (LEP)
& $-0.8$& $-1.5$& $-1.5$& $-1.5$& $-1.5$& $-1.8$& $-1.0$& $-1.3$& $-0.9$& $-1.5$
\\
$M_Z$
& $-1.2$& $-0.4$& $-0.7$& $-0.5$& $-0.3$& $-0.3$& $-1.1$& $-0.6$& $-0.8$& $-0.6$
\\
$\Gamma_Z$
& $0.4$& $-1.4$& $-0.9$& $-1.0$& $-1.4$& $-1.5$& $-0.0$& $-0.7$& $-0.7$& $-1.0$
\\
$\sigma_{h}^{0}$
& $-0.2$& $-0.3$& $2.2$& $-1.0$& $0.8$& $1.0$& $1.4$& $-0.7$& $1.8$& $0.9$
\\
$R^{0}_{e}$
& $1.4$& $1.3$& $0.2$& $1.3$& $0.4$& $0.5$& $0.7$& $1.4$& $0.4$& $0.7$
\\
$R^{0}_{\mu}$
& $1.5$& $1.3$& $1.4$& $-0.3$& $0.0$& $0.1$& $1.5$& $-0.9$& $0.1$& $-0.9$
\\
$R^{0}_{\tau}$
& $-0.3$& $-0.5$& $-0.4$& $-0.4$& $-1.4$& $-0.5$& $-0.3$& $-0.4$& $-1.4$& $-0.4$
\\
$A_{\mathrm{FB}}^{0, e}$
& $-0.7$& $-1.0$& $-0.8$& $-1.0$& $-1.0$& $-1.0$& $-0.7$& $-0.9$& $-0.8$& $-0.9$
\\
$A_{\mathrm{FB}}^{0, \mu}$
& $0.5$& $-0.1$& $0.1$& $0.2$& $-0.0$& $0.0$& $0.4$& $0.4$& $0.4$& $0.4$
\\
$A_{\mathrm{FB}}^{0, \tau}$
& $1.5$& $1.0$& $1.2$& $1.1$& $1.1$& $1.0$& $1.4$& $1.2$& $1.4$& $1.2$
\\
$R^{0}_{b}$
& $0.6$& $0.6$& $0.6$& $0.6$& $0.6$& $0.6$& $0.6$& $0.6$& $0.6$& $0.6$
\\
$R^{0}_{c}$
& $-0.0$& $-0.0$& $-0.0$& $-0.0$& $-0.0$& $-0.0$& $-0.0$& $-0.0$& $-0.0$& $-0.0$
\\
$A_{\mathrm{FB}}^{0, b}$
& $-2.3$& $-3.5$& $-2.6$& $-3.5$& $-3.5$& $-3.4$& $-2.1$& $-3.2$& $-2.6$& $-3.0$
\\
$A_{\mathrm{FB}}^{0, c}$
& $-0.9$& $-1.4$& $-1.0$& $-1.4$& $-1.4$& $-1.3$& $-0.8$& $-1.2$& $-1.0$& $-1.2$
\\
$\mathcal{A}_b$
& $-0.6$& $-0.6$& $-0.6$& $-0.6$& $-0.6$& $-0.6$& $-0.6$& $-0.6$& $-0.6$& $-0.6$
\\
$\mathcal{A}_c$
& $0.1$& $0.0$& $0.0$& $0.0$& $-0.0$& $0.0$& $0.1$& $0.0$& $0.0$& $0.0$
\\
\hline
\end{tabular}
\caption{$IC$ value in each fit and pull for the difference between the measurement and the fit result in units of standard deviation.}
\label{tab:pull}
\vspace*{-20mm}
\end{table}

\section{New physics interpretation}
\label{sec:model}

\begin{table}[t]
\centering
\begin{tabular}{ccc cc cc}
\hline
\crowcolor
& $S_1$ & $\Xi_1$ & 
$E$ & $\Sigma_1$ &
$\mathcal{B}$ & $\mathcal{W}$ \\
Spin &
0 & 0 & 1/2 & 1/2 & 1 & 1
\\
$(\mathrm{SU}(3)_c, \mathrm{SU}(2)_L)_{\mathrm{U}(1)_Y}$ &
$(1,1)_{1}$ &
$(1,3)_{1}$ &
$(1,1)_{-1}$ &
$(1,3)_{-1}$ &
$(1,1)_{0}$ &
$(1,3)_{0}$ \\ [4pt]
\hline
\end{tabular}
\caption{Quantum numbers of the particles considered in this study, where the hypercharge is normalized as $Y=Q-T_{L}^{\prime 3}$.}
\label{tab:particle}
\end{table}

Let us investigate new physics scenarios in light of the $W$-boson mass discrepancy.
The CDF result may indicate extra contributions to the Fermi coupling constant via $(C_{\phi\ell}^{(3)})_{11,22}$ and/or $(C_{\ell\ell})_{1221}$.
We consider simple extensions of the SM, \ie, introduce an extra single scalar, fermion, or vector field at a time.
We assume that their masses are much larger than the Higgs VEV, $v$.
The complete list of the Wilson coefficients of the dimension-six SMEFT operators induced by general single scalar, fermion and vector field is provided at the tree level in Ref.~\cite{deBlas:2017xtg}. 
According to the reference, it is found that there are six types of fields which can contribute to $(C_{\phi\ell}^{(3)})_{11,22}$ and/or $(C_{\ell\ell})_{1221}$.\footnote{
We will not consider a gauge singlet $N \sim (1,1)_{0}$ and an $\mathrm{SU(2)}_L$ adjoint lepton $\Sigma \sim (1,3)_{0}$, because they are likely to generate too large neutrino masses~\cite{Minkowski:1977sc,GellMann:1980vs,Yanagida:1979as,Mohapatra:1979ia,Foot:1988aq}. } 
Their quantum numbers are summarized in Table \ref{tab:particle}.
In this section, we will show by which fields and in which parameter regions the $W$-boson discrepancy is relaxed.\footnote{We will ignore renormalization-group corrections to the SMEFT operators. 
Their effects are expected to be insignificant, because the particles do not carry color charges. 
}

\subsection{Scalar extension}

The $W$-boson mass can be modified by complex scalar fields, $S_1$ and $\Xi_1$, via $(C_{\ell\ell})_{1221}$.
They have Yukawa interactions with the SM (left-handed) leptons as
\begin{align}
 - \mathcal{L}_{\rm int} = 
   (y_{S_1})_{ij} S_1^\dagger \bar\ell_i i\sigma^2 \ell_j^c 
 + (y_{\Xi_1})_{ij} \Xi_1^{a\dagger} \bar\ell_i \sigma^a i\sigma^2 \ell_j^c 
 + {\rm h.c.},
 \label{eq:scalarLagrangian}
\end{align}
where $c$ denotes the charge conjugation. 
Here, $(y_{S_1})_{ij}$ $((y_{\Xi_1})_{ij})$ is anti-symmetric (symmetric) under $i \leftrightarrow j$~\cite{Davies:1990sc}.
Since we are interested in the single-field extension of the SM, $y_{\Xi_1} (y_{S_1}) = 0$ is set when we focus on $S_1$ $(\Xi_1)$. 
The SMEFT operators receive corrections by exchanging the scalar bosons.
If we focus on the above Yukawa interactions, only the Wilson coefficient $C_{\ell\ell}$ among the SMEFT operators is shifted as~\cite{delAguila:2008pw,deBlas:2017xtg}
\begin{align}
 (C_{\ell\ell})_{ijkl} &= 
 \frac{(y_{S_1})_{jl}^*(y_{S_1})_{ik}}{M_{S_1}^2} + \frac{(y_{\Xi_1})_{ki}^*(y_{\Xi_1})_{lj}}{M_{\Xi_1}^2},
 \label{eq:Cllscalar}
\end{align}
at the tree level.
Here $M_{S_1}$ and $M_{\Xi_1}$ are the masses of $S_1$ and $\Xi_1$, respectively.
Although $\Xi_1$ can also have interactions with the SM Higgs boson, they are generically independent of the above Yukawa interaction \eqref{eq:scalarLagrangian} and irrelevant for $G_F$. Hence, we neglect them in the following analysis.

The corrections to the Fermi coupling constant correspond to $(C_{\ell\ell})_{ijkl}$ with $\{i,j,k,l\} = \{1,2,2,1\}$ (see Eq.~\eqref{eq:GF}).
Because of the anti-symmetric or symmetric structure of the Yukawa coupling, the Wilson coefficient becomes
\begin{align}
 (C_{\ell\ell})_{1221} &= 
 -\frac{|(y_{S_1})_{12}|^2}{M_{S_1}^2} + \frac{|(y_{\Xi_1})_{12}|^2}{M_{\Xi_1}^2}. 
\end{align}
From Fig.~\ref{fig:CLL_1221}, it is found that the $W$-boson mass discrepancy favors a positive value for $(C_{\ell\ell})_{1221}$. 
Hence, only $\Xi_1$ can be a source of the anomaly. 
As a result, the couplings and masses are favored to be within the range,
\begin{align}
 0.14 <
 \frac{|(y_{\Xi_1})_{12}|}{M_{\Xi_1}}
 < 0.17 \TeV^{-1}.~~~(68\%)
\end{align}
This result implies that the new physics scale is around $6\text{--}7\TeV$ for $(y_{\Xi_1})_{12} \sim 1$.
The pull of $M_W$ is the same as that for $(C_{\ell\ell})_{1221}$ because the scalar bosons contribute to the EWPO only via this Wilson coefficient.

\subsection{Vector extension}

The $W$-boson mass can be modified by massive vector bosons, $\mathcal{B}$ and $\mathcal{W}$, via $(C_{\ell\ell})_{1221}$.
Here, we do not assume any mechanism to generate the vector boson mass or any UV realization of the model, but consider a low-energy effective framework. 
The vector interactions with the SM (left-handed) leptons are represented as
\begin{align}
 - \mathcal{L}_{\rm int} = 
   (g_{\mathcal{B}})_{ij} \mathcal{B}_\mu \bar \ell_i \gamma^\mu \ell_j
 + \frac{1}{2} (g_{\mathcal{W}})_{ij} \mathcal{W}_\mu^a \bar \ell_i \sigma^a \gamma^\mu \ell_j.
 \label{eq:vectorLagrangian}
\end{align}
Here, $(g_{\mathcal{B,W}})_{ij}$ are hermitian matrices. 
When we focus on $\mathcal{B}$ $(\mathcal{W})$, only $g_{\mathcal{B}}$ $(g_{\mathcal{W}})$ is turned on.
By exchanging the vector bosons, the Wilson coefficient $C_{\ell\ell}$ becomes~\cite{delAguila:2008pw,deBlas:2017xtg}
\begin{align}
 (C_{\ell\ell})_{ijkl} &= 
 - \frac{(g_{\mathcal{B}})_{kl}(g_{\mathcal{B}})_{ij}}{2M_{\mathcal{B}}^2}
 - \frac{(g_{\mathcal{W}})_{kj}(g_{\mathcal{W}})_{il}}{4M_{\mathcal{W}}^2}
 + \frac{(g_{\mathcal{W}})_{kl}(g_{\mathcal{W}})_{ij}}{8M_{\mathcal{W}}^2},
 \label{eq:Cllvector}
\end{align}
at the tree level, where $M_{\mathcal{B}}$ and $M_{\mathcal{W}}$ are the masses of $\mathcal{B}$ and $\mathcal{W}$, respectively. 
Similar to the scalar case, there are no contributions to other SMEFT operators as long as only the interactions \eqref{eq:vectorLagrangian} are considered. 
Although the vector bosons may also have interactions with the SM right-handed leptons, quarks or Higgs boson, they are not always correlated with $(g_{\mathcal{B,W}})_{ij}$. Since they are irrelevant for $G_F$, we neglect them in the following analysis.

The Fermi coupling constant receives corrections from $(C_{\ell\ell})_{ijkl}$ with $\{i,j,k,l\} = \{1,2,2,1\}$ (see Eq.~\eqref{eq:GF}).
Then, the above Wilson coefficient is shown as
\begin{align}
 (C_{\ell\ell})_{1221} &= 
 - \frac{|(g_{\mathcal{B}})_{12}|^2}{2M_{\mathcal{B}}^2}
 - \frac{(g_{\mathcal{W}})_{11}(g_{\mathcal{W}})_{22}}{4M_{\mathcal{W}}^2}
 + \frac{|(g_{\mathcal{W}})_{12}|^2}{8M_{\mathcal{W}}^2}.
\end{align}
The sign of the second term on the right-hand side can be flipped if the product  $(g_{\mathcal{W}})_{11}(g_{\mathcal{W}})_{22}$ is negative.
On the other hand, the first and third terms do not have such a degree of freedom.
Since the $W$-boson mass discrepancy favors a positive value for $(C_{\ell\ell})_{1221}$ as shown in Fig.~\ref{fig:CLL_1221}, only the vector boson $\mathcal{W}$ can be a source of the anomaly. 
Consequently, we obtain
\begin{align}
 0.27 <
 \frac{\mathcal{G}_{\mathcal{W}}}{M_{\mathcal{W}}}
 < 0.35 \TeV^{-1},~~~
 0.38 <
 \frac{(g_{\mathcal{W}})_{12}}{M_{\mathcal{W}}}
 < 0.49 \TeV^{-1},
\end{align}
at the 68\% probability. Here $\mathcal{G}_{\mathcal{W}} = \sqrt{-(g_{\mathcal{W}})_{11}(g_{\mathcal{W}})_{22}}$.
The results imply that the mass of $\mathcal{W}$ is around $2\text{--}4\TeV$ for $g_\mathcal{W}, \mathcal{G}_\mathcal{W} \sim 1$.
Similar to the scalar case, the pull of $M_W$ is the same as that for $(C_{\ell\ell})_{1221}$.

In Eq.~\eqref{eq:Cllvector}, the first and third terms on the right-handed side are understood as neutral-current contributions, while the second term is a charged-current contribution. 
In the presence of $(g_{\mathcal{W}})_{12}$, the muonium-antimuonium oscillation can be induced by exchanging the neutral vector boson, because $|(C_{\ell\ell})_{1212}| = |(C_{\ell\ell})_{1221}|$ due to $(g_{\mathcal{W}})_{12} = (g_{\mathcal{W}})_{21}^*$.
Currently, the experimental constraint is $|(C_{\ell\ell})_{1212}| < 0.1\TeV^{-2}$ at 90\% C.L.~\cite{Willmann:1998gd}.
Thus, the limit is weaker than the parameter required to relax the $W$-boson mass discrepancy.

\begin{figure}[t]
\centering
\includegraphics[scale=0.7]{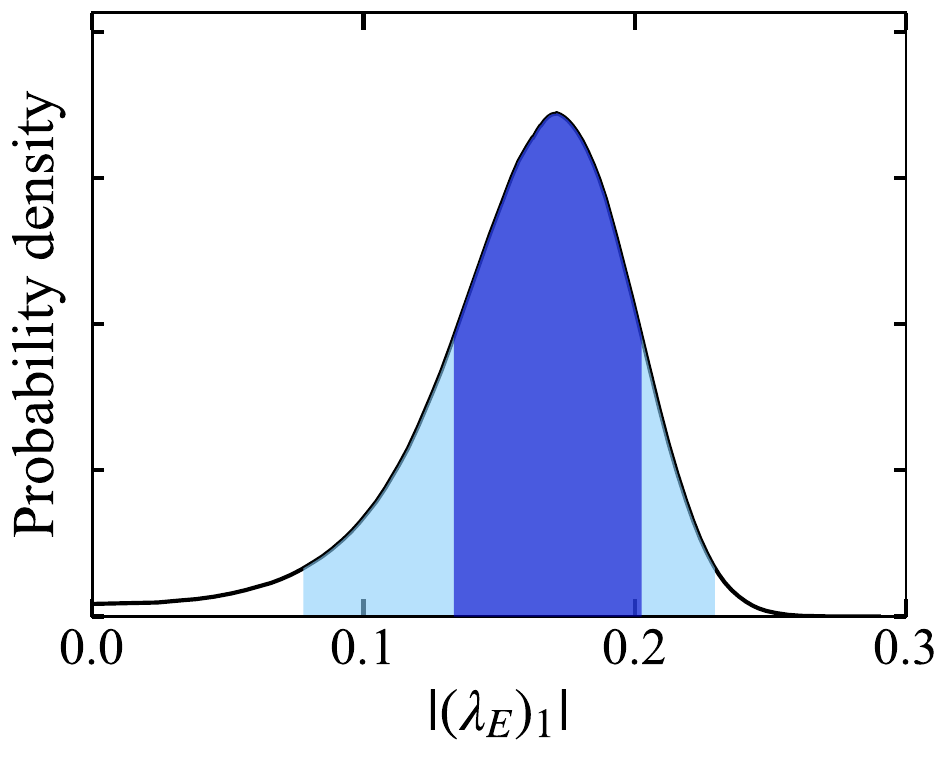}
\hspace{10mm}
\includegraphics[scale=0.7]{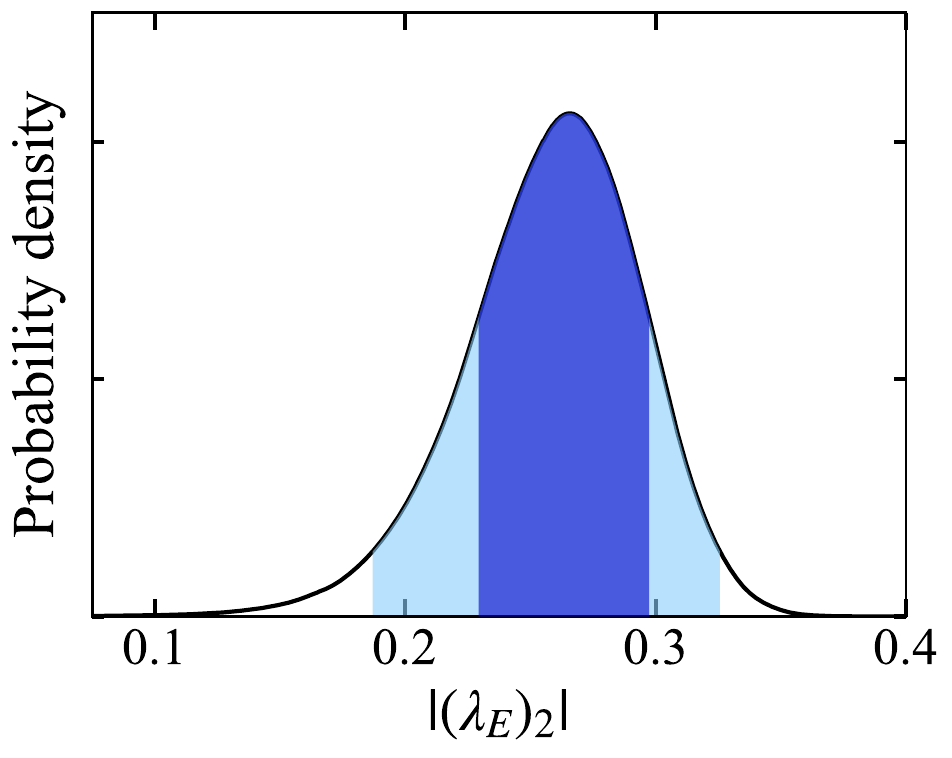}
\\[1mm]
\includegraphics[scale=0.7]{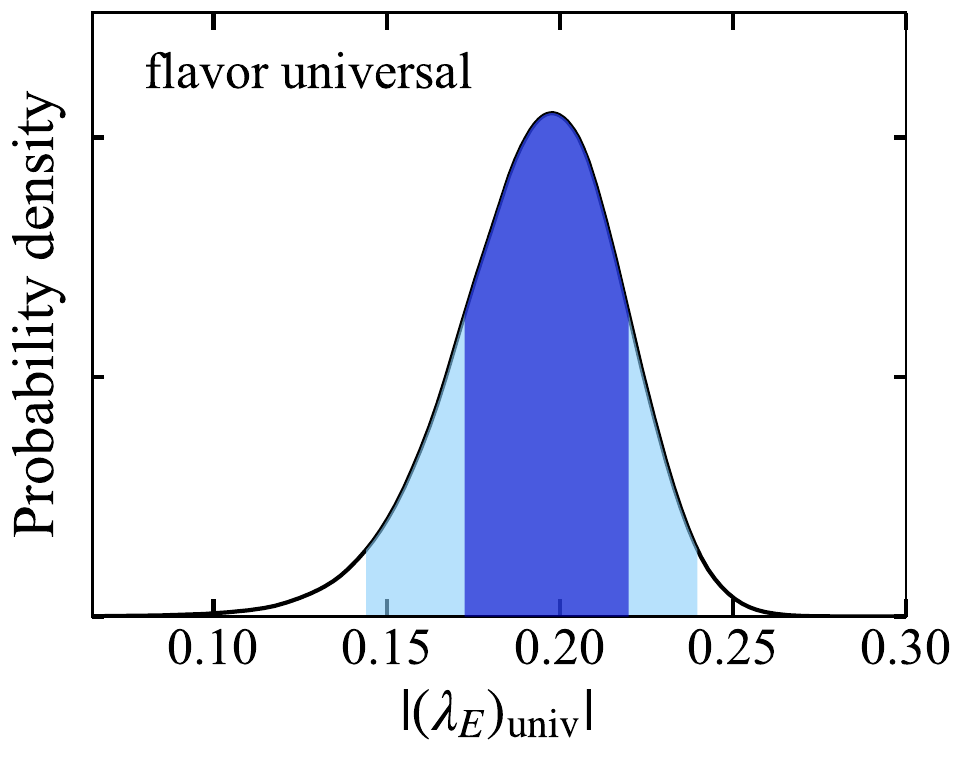}
\hspace{7mm}
\includegraphics[scale=0.7]{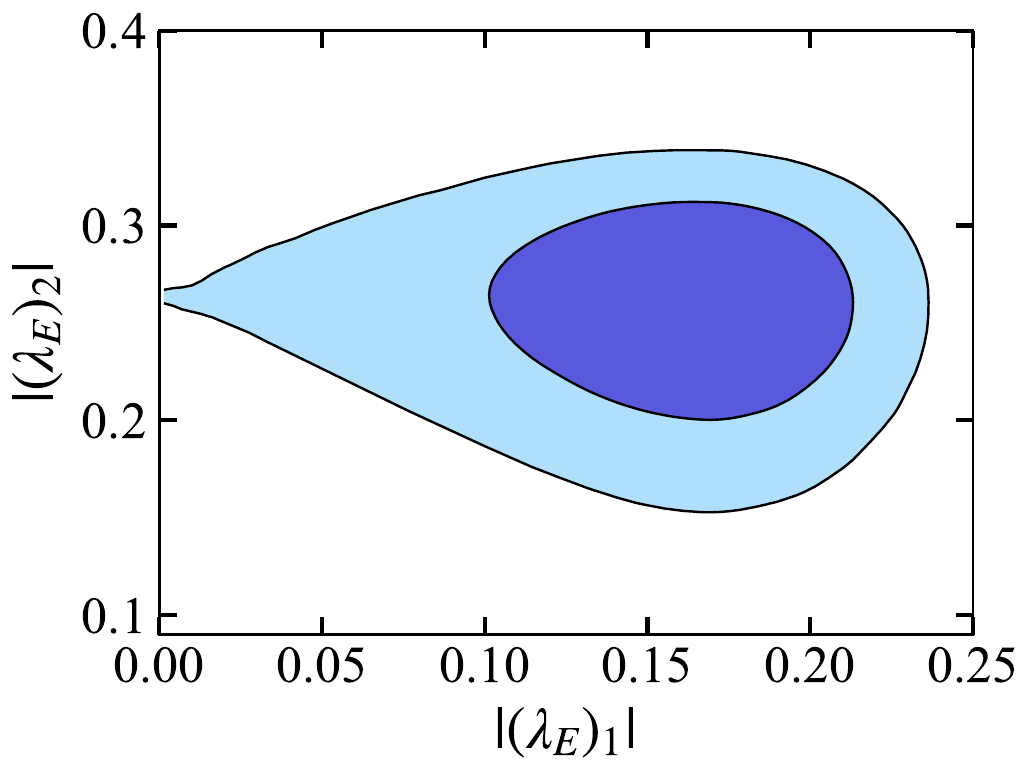}
\caption{Same as Fig.~\ref{fig:CHL3}, but for $\lambda_E$ with $M_E = 1\TeV$.}
\label{fig:lamE}
\end{figure}

\subsection{Fermion extension}

As the third scenario, let us consider extra leptons, $E$ and $\Sigma_1$.
The Yukawa interactions with the SM (left-handed) leptons are shown as
\begin{align}
 -\mathcal{L}_{\rm int} &= 
   (\lambda_E)_{i}\, \bar E_R \phi^\dagger \ell_i
 + \frac{1}{2} (\lambda_{\Sigma_1})_{i}\, \bar \Sigma_{1R}^a \phi^\dagger \sigma^a \ell_i
 + {\rm h.c.}
 \label{eq:Yukawa}
\end{align}
In addition, there are vectorlike mass terms, $-\mathcal{L}_{\rm mass} = M'_E\, \bar E_L E_R + M'_{\Sigma_1}\, \bar \Sigma_{1L}^a \Sigma_{1R}^a + {\rm h.c}$, with $M'_E, M'_{\Sigma_1} \gg v$. 
Hereafter, let us denote the mass eigenvalues of the extra leptons as $M_E (\simeq M'_E)$ and $M_{\Sigma_1} (\simeq M'_{\Sigma_1})$.
Also, only $\lambda_E$ $(\lambda_{\Sigma_1})$ is turned on in the $E$ $(\Sigma_1)$ scenario.
Then, the Wilson coefficients are obtained as~\cite{delAguila:2008pw,deBlas:2017xtg}
\begin{align}
 (C_{e\phi})_{ij} &= (y_\ell)_{jk}^* \left[
   \frac{(\lambda_E)_{k}(\lambda_E)_{i}^*}{2M_E^2} 
 + \frac{(\lambda_{\Sigma_1})_{k}(\lambda_{\Sigma_1})_{i}^*}{8M_{\Sigma_1}^2}
 \right],
 \label{eq:CEH} \\
 (C_{\phi\ell}^{(1)})_{ij} &= 
 - \frac{(\lambda_E)_{j}(\lambda_E)_{i}^*}{4M_E^2}
 - \frac{3(\lambda_{\Sigma_1})_{j}(\lambda_{\Sigma_1})_{i}^*}{16M_{\Sigma_1}^2}, 
 \label{eq:CHE1}\\
 (C_{\phi\ell}^{(3)})_{ij} &= 
 - \frac{(\lambda_E)_{j}(\lambda_E)_{i}^*}{4M_E^2}
 + \frac{(\lambda_{\Sigma_1})_{j}(\lambda_{\Sigma_1})_{i}^*}{16M_{\Sigma_1}^2}, 
 \label{eq:CHE3}
\end{align}
at the tree level.\footnote{
Lepton flavors are violated generally if the extra leptons couple to multiple SM leptons, \eg, the electron and muon,  simultaneously. 
Such violations are avoided by assuming that each extra lepton couples to only one of the electron, muon and tau-flavor leptons. 
Hence, $(C_{e\phi})_{ij} = (C_{\phi\ell}^{(1)})_{ij} = (C_{\phi\ell}^{(3)})_{ij} = 0$ for $i \neq j$. 
Then, multiple extra leptons are introduced when multiple $(\lambda_E)_{i}$ are turned on. 
In the following discussions, we assume that their masses are common. 
}
Here $(y_\ell)_{ij}$ is the lepton Yukawa coupling. 
The $W$-boson mass receives corrections from $(C_{\phi\ell}^{(3)})_{11,22}$ via the Fermi coupling constant, while the $Z$-boson observables are additionally affected by $(C_{\phi\ell}^{(1,3)})_{ii}$ directly, and the $W$-boson partial decay width $\Gamma(W^+ \to \ell_i^+ \nu_{\ell i})$ are by $(C_{\phi\ell}^{(3)})_{ii}$ (see Sec.~\ref{sec:EWPO}).

For the Fermi coupling constant, the Wilson coefficient in Eq.~\eqref{eq:CHE3} is shown as
\begin{align}
 (C_{\phi\ell}^{(3)})_{11,22} &= 
 - \frac{|(\lambda_E)_{1,2}|^2}{4M_E^2}
 + \frac{|(\lambda_{\Sigma_1})_{1,2}|^2}{16M_{\Sigma_1}^2}. 
\end{align}
Since the contribution from the extra lepton $E$ ($\Sigma_1$) is negative (positive), it is found from Fig.~\ref{fig:CHL3} that $E$ can be a source of the $W$-boson mass discrepancy.
Therefore, we focus on $E$ in the following.

The extra lepton $E$ induces the SMEFT operators 
$\mathcal{O}_{e\phi}$, $\mathcal{O}_{\phi\ell}^{(1)}$ and $\mathcal{O}_{\phi\ell}^{(3)}$, 
where the EWPO are insensitive to the first one. 
With including all the contributions, we obtain the EWPO fit of the model parameters with a lepton-flavor dependent or universal assumption. 
The results are shown in Fig.~\ref{fig:lamE}.
In the top two plots, either $(\lambda_E)_{1}$ or $(\lambda_E)_{2}$ is switched on, while in the bottom-left plot, we assume $(\lambda_E)_{1} = (\lambda_E)_{2} = (\lambda_E)_{3} \equiv (\lambda_E)_{\mathrm{univ}}$ and in the bottom-right plot, $(\lambda_E)_{1}$ and $(\lambda_E)_{2}$ are fitted simultaneously with $(\lambda_E)_{3} = 0$.

From the figure, it is found that the muonic interaction $(\lambda_E)_{2}$ is required to be larger than $(\lambda_E)_{1}$; the $W$-boson mass discrepancy implies that the extra lepton $E$ may exist in $5\text{--}7\TeV$ for $(\lambda_E)_{1} \sim 1$ and around $3\text{--}4\TeV$ for $(\lambda_E)_{2} \sim 1$.
For the universal case, the extra lepton mass is favored to be $5\text{--}6\TeV$ for $(\lambda_E)_{\mathrm{univ}} \sim 1$.
From Table \ref{tab:pull}, we also found that the pull of $M_W$ is slightly worse than the result of $C_{\phi\ell}^{(3)}$.
This is because of the additional contributions to the EWPO via $C_{\phi\ell}^{(1)}$.
However, the degradation is milder when $(\lambda_E)_{1}$ and $(\lambda_E)_{2}$ are fitted simultaneously. 

\begin{figure}[t]
\centering
\includegraphics[scale=0.7]{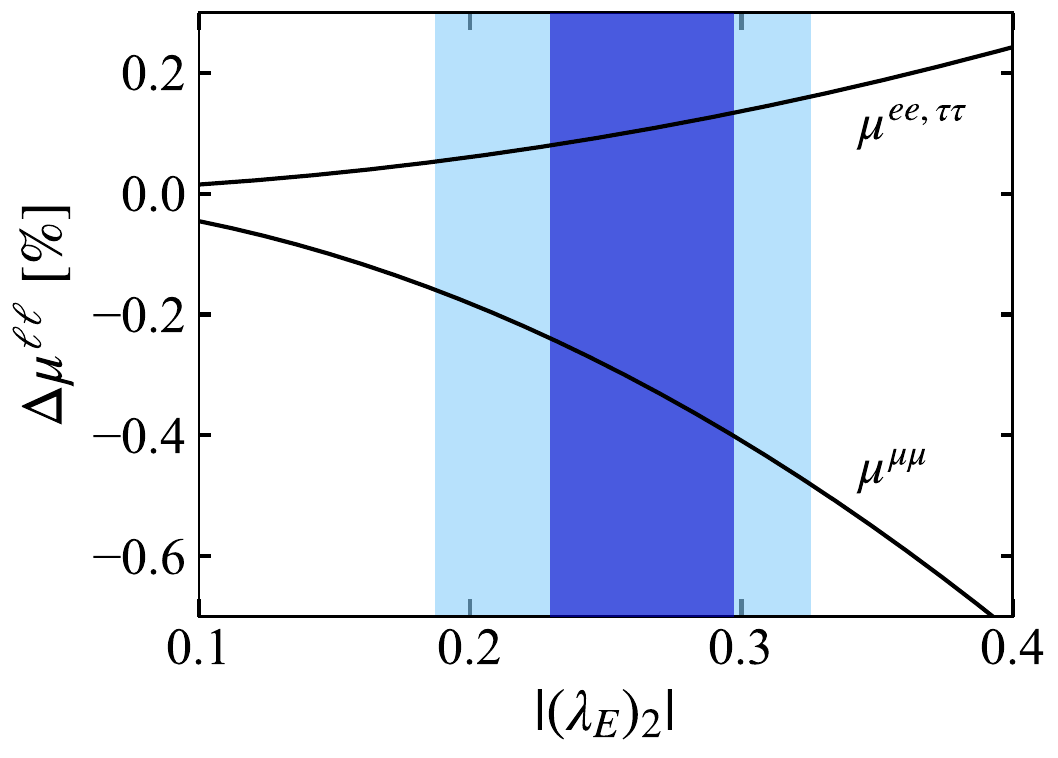}
\hspace{5mm}
\includegraphics[scale=0.7]{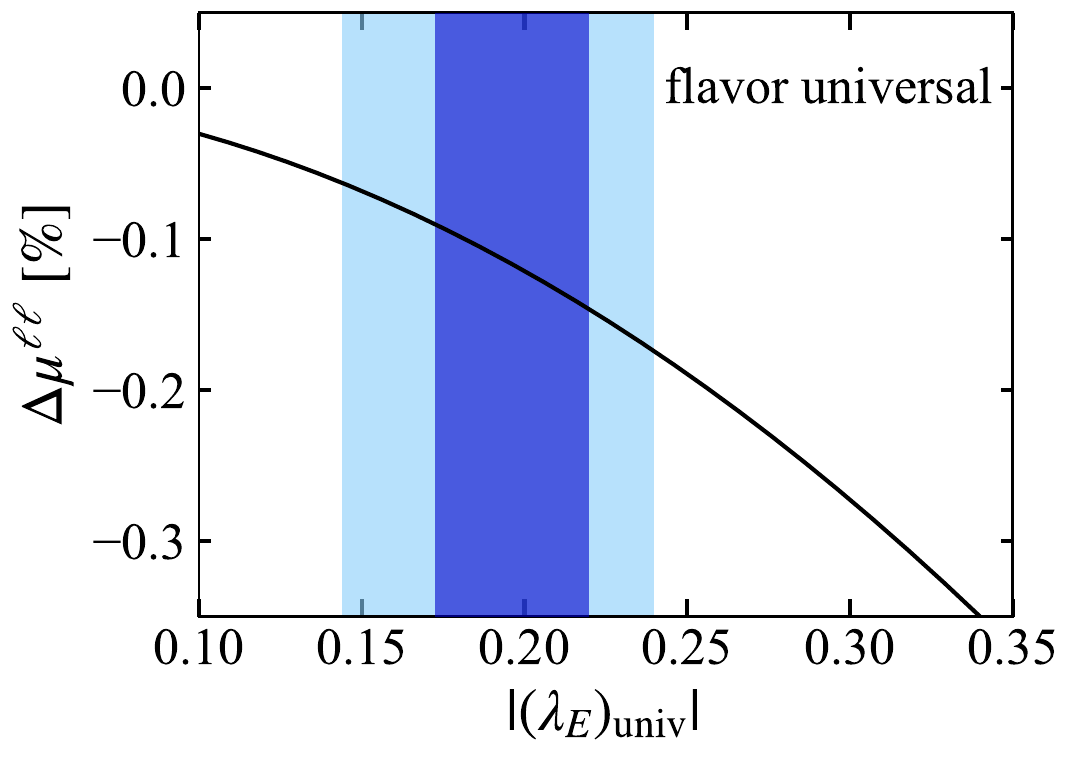}
\caption{Signal strength of the Higgs boson decays for $\lambda_E$ with $M_E = 1\TeV$.
Only $(\lambda_{E})_{2}$ is switched on in the left plot, while the coupling is universal in the right. 
The darker (lighter) region corresponds to the 68\% (95\%) probability provided in Fig.~\ref{fig:lamE}.
}
\label{fig:higgs}
\end{figure}

The extra leptons can also affect the Higgs interactions via the SMEFT operator $\mathcal{O}_{e\phi}$.
The lepton Yukawa coupling is shifted as
\begin{align}
 y_{\ell i} 
 = \sqrt{2}\,\frac{m_{\ell i}}{v} + \frac{1}{2}\, (\widehat{C}_{e\phi})_{ii},
 = (y_{ei})_{\rm SM} \left[ 1 - \frac{1}{2}\, \delta_{G_F} \right]
 + \frac{1}{2}\, (\widehat{C}_{e\phi})_{ii},
 \label{eq:leptonYukawa}
\end{align}
after the electroweak symmetry breaking.
Then, the signal strength of the Higgs decay into lepton pair is modified from the SM prediction as
\begin{align}
 \mu^{\ell_i\ell_i} 
 \equiv \frac{\Gamma(h \to \ell_i\ell_i)}{\Gamma(h \to \ell_i\ell_i)_{\rm SM}}
 = \left| 
 1 - \frac{1}{2}\, \delta_{G_F}
 - \frac{1}{(y_{\ell i})_{\rm SM}}\, (\widehat{C}_{e\phi})_{ii}
 \right|^2.
\end{align}
In Fig.~\ref{fig:higgs}, the theoretical predictions of $\mu^{\mu\mu}$ and $\mu^{\ell\ell}$ are shown for the cases when only $(\lambda_{E})_{2}$ is switched on (left) and the universal coupling $(\lambda_{E})_{\mathrm{univ}}$ is varied (right). 
Here $\mu^{ee} = \mu^{\mu\mu} = \mu^{\tau\tau}$ in the right plot.
The vertical axis is the deviation of the signal strength from the SM prediction, $\Delta \mu^{\ell_i\ell_i} = \mu^{\ell_i\ell_i} - 1$.
It is found that the signal strengths are changed by $\mathcal{O}(0.1)\%$ by the contributions of $E$. 
The effect is sufficiently weaker than the current experimental sensitivities; the experimental results are $\mu^{\mu\mu} = 1.2\pm0.6$~\cite{ATLAS:2020fzp},
$\mu^{\tau\tau} = 1.14\pm0.32$~\cite{ATLAS:2019nkf} from ATLAS, and 
$\mu^{\mu\mu} = 1.19^{+0.40}_{-0.39}{\rm (stat)}^{+0.15}_{-0.14}{\rm (syst)}$~\cite{CMS:2020xwi},
$\mu^{\tau\tau} = 1.02^{+0.26}_{-0.24}$~\cite{CMS:2018uag} from CMS.
In future, the HL-LHC experiment may achieve $\delta \mu^{\mu\mu}/\mu^{\mu\mu} = 9$\% and $\delta \mu^{\tau\tau}/\mu^{\tau\tau} = 4\%$ at $\sqrt{s}=14\TeV$ with the integrated luminosity $\mathcal{L}=6$\,ab$^{-1}$, and the sensitivities could reach $\delta \mu^{\mu\mu}/\mu^{\mu\mu} = 0.8\%$ and $\delta \mu^{\tau\tau}/\mu^{\tau\tau} = 0.9\%$ at FCC-ee/eh/hh~\cite{deBlas:2019rxi}.
Therefore, further improvement is necessary to detect the effects of the extra lepton.

\section{Conclusions and discussion}
\label{sec:conclusion}

Motivated by the tensions reported in the updated measurement of the $W$-boson mass by the CDF collaboration, we studied the new physics scenarios that affect $G_F$. 
We investigated single-field extensions of the SM, \ie, introducing one of the extra scalars ($S_1$ and $\Xi_1$), leptons ($E$ and $\Sigma_1$), and vector fields ($\mathcal{B}$ and $\mathcal{W}$).
It was found that the models with $\Xi_1$, $E$ and $\mathcal{W}$ can relax the discrepancy if the new particles exist in multi-TeV scales when the new coupling to the SM leptons is of order unity. 
In particular, the scalars and vectors affect $M_W$ via the lepton four-Fermi operator $\mathcal{O}_{\ell\ell}$; its Wilson coefficient is favored to be $\sim 0.02\text{--}0.03\TeV^{-2}$.
On the other hand, the extra leptons contribute to the Fermi coupling constant via $\mathcal{O}_{\phi\ell}^{(3)}$, and 
the $M_W$ discrepancy implies $(C_{\phi\ell}^{(3)})_{11}$ and/or $(C_{\phi\ell}^{(3)})_{22}$ around $-(0.01\text{--}0.03)\TeV^{-2}$,
leading to $(\lambda_E)_{1}$ and/or $(\lambda_E)_{2}$ around $0.1\text{--}0.3$ for $M_E = 1\TeV$.
We also evaluated the pull of EWPO including $M_W$ in each fit. 
Although any of the scenarios cannot explain the discrepancy perfectly, the tension is shown to be relaxed.
In particular, the lepton-flavor universal contribution of $C_{\phi\ell}^{(3)}$ provides the best result.

If new scalar and vector bosons exist in multi-TeV scales, they could be detected via resonance searches of $pp \to S,V \to \ell\ell$, if they couple to the SM quarks. 
However, such interactions are not necessary for the $W$-boson discrepancy.
The required mass scales decrease if the couplings to the SM leptons are weaker. 
Then, the new particles could be produced via the gauge interactions (except for $\mathcal{B}$). 
They decay into the SM leptons and might be discovered in searching for multi-lepton signatures. 
Such a study has been performed by ATLAS with the full Run 2 dataset of $139\,{\rm fb}^{-1}$~\cite{ATLAS:2021wob}, though the result has not been applied to the current setup. 
On the other hand, the extra fermions decay into the SM bosons as well as the SM leptons. 
ATLAS and CMS have performed such studies with the full Run 2 dataset~\cite{ATLAS:2022yhd,CMS:2022nty} and provided limits on the extra-lepton mass as $M \gtrsim 1\TeV$.
The sensitivity might be able to reach multi-TeV scales in future experiments such as HL-LHC or higher energy colliders~\cite{Bhattiprolu:2019vdu}.
Thus, the extra-lepton model could be tested in future. 

Let us comment on connections with other anomalies. 
Currently, the experimental results of the anomalous magnetic moment of muon~\cite{Muong-2:2002wip,Muong-2:2004fok,Muong-2:2006rrc,Muong-2:2021ojo} show a $4.2\sigma$ discrepancy from the SM prediction~\cite{Aoyama:2020ynm}.
This tension can be solved by introducing the vectorlike leptons at the one-loop level~\cite{Czarnecki:2001pv,Kannike:2011ng,Dermisek:2013gta,Freitas:2014pua,Megias:2017dzd,Poh:2017tfo,Kowalska:2017iqv,Raby:2017igl,Calibbi:2018rzv,Crivellin:2018qmi,Kawamura:2019rth,Kawamura:2019hxp,Endo:2020tkb,Crivellin:2020ebi,Athron:2021iuf,Djouadi:2021wvb}.  
In fact, the extra contributions can be as large as $\mathcal{O}(10^{-9})$ even for $M_i = \mathcal{O}(1)\TeV$ by introducing another extra lepton, $\Delta_{1} \sim (1,2)_{-\frac{1}{2}}$ or $\Delta_{3} \sim (1,2)_{-\frac{3}{2}}$, in addition to $E$, because interactions among the extra leptons can enhance a chirality flip. 
Note that $\Delta_{1}$ and $\Delta_{3}$ do not contribute to $G_F$, while they affect EWPO via the SMEFT operator $(\phi^\dagger i \olrarrow{D_\mu} \phi) (\bar e_i \gamma^\mu e_j)$.
Hence, further analyses are required for detailed studies.

\section*{Note added}
When we were finalizing this paper, Ref.~\cite{Bagnaschi:2022whn} appeared on arXiv, which discusses single field extensions of the SM. In contrast to that paper, we investigated here in details the scenarios where the $W$-boson mass anomaly is relaxed by the new physics affecting the Fermi constant.  
In particular, we studied the effects of the flavor non-universal couplings in  addition to the flavor universal ones. 
Furthermore, we considered the scalar field $\Xi_1$, which was not studied in Ref.~\cite{Bagnaschi:2022whn}.

\section*{Acknowledgements}
The authors thank T.~Felkl for valuable comments.
This work is supported in part by the Japan Society for the Promotion of Science (JSPS) Grant-in-Aid for Scientific Research on Innovative Areas (No.~21H00086 [ME]) and Scientific Research B (No.~21H01086 [ME]).

\bibliographystyle{utphys28mod}
\bibliography{references}

\providecommand{\href}[2]{#2}\begingroup\raggedright\begin{thebibliography}{10}

\bibitem{CDF:2022hxs}
{\bfseries CDF} Collaboration, {\em {High-precision measurement of the $W$
  boson mass with the CDF II detector},}
  \href{https://dx.doi.org/10.1126/science.abk1781}{Science {\bfseries 376}
  (2022) 170--176}.

\bibitem{deBlas:2022hdk}
J.~de~Blas, M.~Pierini, L.~Reina, and L.~Silvestrini, {\em {Impact of the
  recent measurements of the top-quark and W-boson masses on electroweak
  precision fits}.} {\ttfamily
  \href{https://arxiv.org/abs/2204.04204}{arXiv:2204.04204}}.

\bibitem{ATLAS:2017rzl}
{\bfseries ATLAS} Collaboration, {\em {Measurement of the $W$-boson mass in pp
  collisions at $\sqrt{s}=7$ TeV with the ATLAS detector},}
  \href{https://dx.doi.org/10.1140/epjc/s10052-017-5475-4}{Eur.\  Phys.\  J.\
  C {\bfseries 78} (2018) 110} {\ttfamily
  [\href{https://arxiv.org/abs/1701.07240}{arXiv:1701.07240}]}. [Erratum:
  Eur.Phys.J.C 78, 898 (2018)].

\bibitem{LHCb:2021bjt}
{\bfseries LHCb} Collaboration, {\em {Measurement of the W boson mass},}
  \href{https://dx.doi.org/10.1007/JHEP01(2022)036}{JHEP {\bfseries 01} (2022)
  036} {\ttfamily [\href{https://arxiv.org/abs/2109.01113}{arXiv:2109.01113}]}.

\bibitem{Fan:2022dck}
Y.-Z.~Fan, T.-P.~Tang, Y.-L.~S.~Tsai, and L.~Wu, {\em {Inert Higgs Dark Matter
  for New CDF W-boson Mass and Detection Prospects}.} {\ttfamily
  \href{https://arxiv.org/abs/2204.03693}{arXiv:2204.03693}}.

\bibitem{Zhu:2022tpr}
C.-R.~Zhu, {\em et al.}, {\em {GeV antiproton/gamma-ray excesses and the
  $W$-boson mass anomaly: three faces of $\sim 60-70$ GeV dark matter
  particle?}} {\ttfamily
  \href{https://arxiv.org/abs/2204.03767}{arXiv:2204.03767}}.

\bibitem{Lu:2022bgw}
C.-T.~Lu, L.~Wu, Y.~Wu, and B.~Zhu, {\em {Electroweak Precision Fit and New
  Physics in light of $W$ Boson Mass}.} {\ttfamily
  \href{https://arxiv.org/abs/2204.03796}{arXiv:2204.03796}}.

\bibitem{Athron:2022qpo}
P.~Athron, {\em et al.}, {\em {The $W$ boson Mass and Muon $g-2$: Hadronic
  Uncertainties or New Physics?}} {\ttfamily
  \href{https://arxiv.org/abs/2204.03996}{arXiv:2204.03996}}.

\bibitem{Yuan:2022cpw}
G.-W.~Yuan, L.~Zu, L.~Feng, and Y.-F.~Cai, {\em {$W$-boson mass anomaly:
  probing the models of axion-like particle, dark photon and Chameleon dark
  energy}.} {\ttfamily
  \href{https://arxiv.org/abs/2204.04183}{arXiv:2204.04183}}.

\bibitem{Strumia:2022qkt}
A.~Strumia, {\em {Interpreting electroweak precision data including the
  $W$-mass CDF anomaly}.} {\ttfamily
  \href{https://arxiv.org/abs/2204.04191}{arXiv:2204.04191}}.

\bibitem{Yang:2022gvz}
J.~M.~Yang and Y.~Zhang, {\em {Low energy SUSY confronted with new measurements
  of W-boson mass and muon g-2}.} {\ttfamily
  \href{https://arxiv.org/abs/2204.04202}{arXiv:2204.04202}}.

\bibitem{Du:2022pbp}
X.~K.~Du, Z.~Li, F.~Wang, and Y.~K.~Zhang, {\em {Explaining The Muon $g-2$
  Anomaly and New CDFII W-Boson Mass in the Framework of ExtraOrdinary Gauge
  Mediation}.} {\ttfamily
  \href{https://arxiv.org/abs/2204.04286}{arXiv:2204.04286}}.

\bibitem{Tang:2022pxh}
T.-P.~Tang, M.~Abdughani, L.~Feng, Y.-L.~S.~Tsai, and Y.-Z.~Fan, {\em {NMSSM
  neutralino dark matter for $W$-boson mass and muon $g-2$ and the promising
  prospect of direct detection}.} {\ttfamily
  \href{https://arxiv.org/abs/2204.04356}{arXiv:2204.04356}}.

\bibitem{Cacciapaglia:2022xih}
G.~Cacciapaglia and F.~Sannino, {\em {The W boson mass weighs in on the
  non-standard Higgs}.} {\ttfamily
  \href{https://arxiv.org/abs/2204.04514}{arXiv:2204.04514}}.

\bibitem{Blennow:2022yfm}
M.~Blennow, P.~Coloma, E.~Fern\'andez-Mart\'\i{}nez, and M.~Gonz\'alez-L\'opez,
  {\em {Right-handed neutrinos and the CDF II anomaly}.} {\ttfamily
  \href{https://arxiv.org/abs/2204.04559}{arXiv:2204.04559}}.

\bibitem{Arias-Aragon:2022ats}
F.~Arias-Arag\'on, E.~Fern\'andez-Mart\'\i{}nez, M.~Gonz\'alez-L\'opez, and
  L.~Merlo, {\em {Dynamical Minimal Flavour Violating Inverse Seesaw}.}
  {\ttfamily \href{https://arxiv.org/abs/2204.04672}{arXiv:2204.04672}}.

\bibitem{Zhu:2022scj}
B.-Y.~Zhu, S.~Li, J.-G.~Cheng, R.-L.~Li, and Y.-F.~Liang, {\em {Using gamma-ray
  observation of dwarf spheroidal galaxy to test a dark matter model that can
  interpret the W-boson mass anomaly}.} {\ttfamily
  \href{https://arxiv.org/abs/2204.04688}{arXiv:2204.04688}}.

\bibitem{Sakurai:2022hwh}
K.~Sakurai, F.~Takahashi, and W.~Yin, {\em {Singlet extensions and W boson mass
  in the light of the CDF II result}.} {\ttfamily
  \href{https://arxiv.org/abs/2204.04770}{arXiv:2204.04770}}.

\bibitem{Fan:2022yly}
J.~Fan, L.~Li, T.~Liu, and K.-F.~Lyu, {\em {$W$-Boson Mass, Electroweak
  Precision Tests and SMEFT}.} {\ttfamily
  \href{https://arxiv.org/abs/2204.04805}{arXiv:2204.04805}}.

\bibitem{Liu:2022jdq}
X.~Liu, S.-Y.~Guo, B.~Zhu, and Y.~Li, {\em {Unifying gravitational waves with
  $W$ boson, FIMP dark matter, and Majorana Seesaw mechanism}.} {\ttfamily
  \href{https://arxiv.org/abs/2204.04834}{arXiv:2204.04834}}.

\bibitem{Lee:2022nqz}
H.~M.~Lee and K.~Yamashita, {\em {A Model of Vector-like Leptons for the Muon
  $g-2$ and the $W$ Boson Mass}.} {\ttfamily
  \href{https://arxiv.org/abs/2204.05024}{arXiv:2204.05024}}.

\bibitem{Cheng:2022jyi}
Y.~Cheng, X.-G.~He, Z.-L.~Huang, and M.-W.~Li, {\em {Type-II Seesaw Triplet
  Scalar and Its VEV Effects on Neutrino Trident Scattering and W mass}.}
  {\ttfamily \href{https://arxiv.org/abs/2204.05031}{arXiv:2204.05031}}.

\bibitem{Song:2022xts}
H.~Song, W.~Su, and M.~Zhang, {\em {Electroweak Phase Transition in 2HDM under
  Higgs, Z-pole, and W precision measurements}.} {\ttfamily
  \href{https://arxiv.org/abs/2204.05085}{arXiv:2204.05085}}.

\bibitem{Bagnaschi:2022whn}
E.~Bagnaschi, {\em et al.}, {\em {SMEFT Analysis of $m_{W}$}.} {\ttfamily
  \href{https://arxiv.org/abs/2204.05260}{arXiv:2204.05260}}.

\bibitem{Paul:2022dds}
A.~Paul and M.~Valli, {\em {Violation of custodial symmetry from W-boson mass
  measurements}.} {\ttfamily
  \href{https://arxiv.org/abs/2204.05267}{arXiv:2204.05267}}.

\bibitem{Bahl:2022xzi}
H.~Bahl, J.~Braathen, and G.~Weiglein, {\em {New physics effects on the
  $W$-boson mass from a doublet extension of the SM Higgs sector}.} {\ttfamily
  \href{https://arxiv.org/abs/2204.05269}{arXiv:2204.05269}}.

\bibitem{Asadi:2022xiy}
P.~Asadi, C.~Cesarotti, K.~Fraser, S.~Homiller, and A.~Parikh, {\em {Oblique
  Lessons from the $W$ Mass Measurement at CDF II}.} {\ttfamily
  \href{https://arxiv.org/abs/2204.05283}{arXiv:2204.05283}}.

\bibitem{DiLuzio:2022xns}
L.~Di~Luzio, R.~Gr\"ober, and P.~Paradisi, {\em {Higgs physics confronts the
  $M_W$ anomaly}.} {\ttfamily
  \href{https://arxiv.org/abs/2204.05284}{arXiv:2204.05284}}.

\bibitem{Athron:2022isz}
P.~Athron, {\em et al.}, {\em {Precise calculation of the W boson pole mass
  beyond the Standard Model with FlexibleSUSY}.} {\ttfamily
  \href{https://arxiv.org/abs/2204.05285}{arXiv:2204.05285}}.

\bibitem{Gu:2022htv}
J.~Gu, Z.~Liu, T.~Ma, and J.~Shu, {\em {Speculations on the W-Mass Measurement
  at CDF}.} {\ttfamily
  \href{https://arxiv.org/abs/2204.05296}{arXiv:2204.05296}}.

\bibitem{Heckman:2022the}
J.~J.~Heckman, {\em {Extra $W$-Boson Mass from a D3-Brane}.} {\ttfamily
  \href{https://arxiv.org/abs/2204.05302}{arXiv:2204.05302}}.

\bibitem{Babu:2022pdn}
K.~S.~Babu, S.~Jana, and V.~P.~K., {\em {Correlating $W$-Boson Mass Shift with
  Muon $g-2$ in the 2HDM}.} {\ttfamily
  \href{https://arxiv.org/abs/2204.05303}{arXiv:2204.05303}}.

\bibitem{Grzadkowski:2010es}
B.~Grzadkowski, M.~Iskrzynski, M.~Misiak, and J.~Rosiek, {\em {Dimension-Six
  Terms in the Standard Model Lagrangian},}
  \href{https://dx.doi.org/10.1007/JHEP10(2010)085}{JHEP {\bfseries 10} (2010)
  085} {\ttfamily [\href{https://arxiv.org/abs/1008.4884}{arXiv:1008.4884}]}.

\bibitem{Ciuchini:2013pca}
M.~Ciuchini, E.~Franco, S.~Mishima, and L.~Silvestrini, {\em {Electroweak
  Precision Observables, New Physics and the Nature of a 126 GeV Higgs Boson},}
  \href{https://dx.doi.org/10.1007/JHEP08(2013)106}{JHEP {\bfseries 08} (2013)
  106} {\ttfamily [\href{https://arxiv.org/abs/1306.4644}{arXiv:1306.4644}]}.

\bibitem{ALEPH:2005ab}
{\bfseries ALEPH, DELPHI, L3, OPAL, SLD, LEP Electroweak Working Group, SLD
  Electroweak Group, SLD Heavy Flavour Group} Collaboration, {\em {Precision
  electroweak measurements on the $Z$ resonance},}
  \href{https://dx.doi.org/10.1016/j.physrep.2005.12.006}{Phys.\ \ Rept.\
  {\bfseries 427} (2006) 257--454} {\ttfamily
  [\href{https://arxiv.org/abs/hep-ex/0509008}{hep-ex/0509008}]}.

\bibitem{Schael:2013ita}
{\bfseries ALEPH, DELPHI, L3, OPAL, LEP Electroweak} Collaboration, {\em
  {Electroweak Measurements in Electron-Positron Collisions at W-Boson-Pair
  Energies at LEP},}
  \href{https://dx.doi.org/10.1016/j.physrep.2013.07.004}{Phys.\ \ Rept.\
  {\bfseries 532} (2013) 119--244} {\ttfamily
  [\href{https://arxiv.org/abs/1302.3415}{arXiv:1302.3415}]}.

\bibitem{Janot:2019oyi}
P.~Janot and S.~Jadach, {\em {Improved Bhabha cross section at LEP and the
  number of light neutrino species},}
  \href{https://dx.doi.org/10.1016/j.physletb.2020.135319}{Phys.\  Lett.\
  {\bfseries B803} (2020) 135319}
{\ttfamily [\href{https://arxiv.org/abs/1912.02067}{arXiv:1912.02067}]}.

\bibitem{deBlas:2019okz}
J.~de~Blas {\em et~al.}, {\em {HEPfit: a Code for the Combination of Indirect
  and Direct Constraints on High Energy Physics Models}.}
{\ttfamily \href{https://arxiv.org/abs/1910.14012}{arXiv:1910.14012}}.

\bibitem{Caldwell:2008fw}
A.~Caldwell, D.~Kollar, and K.~Kroninger, {\em {BAT: The Bayesian Analysis
  Toolkit},} \href{https://dx.doi.org/10.1016/j.cpc.2009.06.026}{Comput.\
  Phys.\  Commun.\  {\bfseries 180} (2009) 2197--2209}
{\ttfamily [\href{https://arxiv.org/abs/0808.2552}{arXiv:0808.2552}]}.

\bibitem{Awramik:2003rn}
M.~Awramik, M.~Czakon, A.~Freitas, and G.~Weiglein, {\em {Precise prediction
  for the W boson mass in the standard model},}
  \href{https://dx.doi.org/10.1103/PhysRevD.69.053006}{Phys.\ \ Rev.\ \ D
  {\bfseries 69} (2004) 053006} {\ttfamily
  [\href{https://arxiv.org/abs/hep-ph/0311148}{hep-ph/0311148}]}.

\bibitem{Awramik:2006uz}
M.~Awramik, M.~Czakon, and A.~Freitas, {\em {Electroweak two-loop corrections
  to the effective weak mixing angle},}
  \href{https://dx.doi.org/10.1088/1126-6708/2006/11/048}{JHEP {\bfseries 11}
  (2006) 048} {\ttfamily
  [\href{https://arxiv.org/abs/hep-ph/0608099}{hep-ph/0608099}]}.

\bibitem{Dubovyk:2019szj}
I.~Dubovyk, A.~Freitas, J.~Gluza, T.~Riemann, and J.~Usovitsch, {\em
  {Electroweak pseudo-observables and Z-boson form factors at two-loop
  accuracy},} \href{https://dx.doi.org/10.1007/JHEP08(2019)113}{JHEP {\bfseries
  08} (2019) 113} {\ttfamily
  [\href{https://arxiv.org/abs/1906.08815}{arXiv:1906.08815}]}.

\bibitem{Bardin:1986fi}
D.~Y.~Bardin, S.~Riemann, and T.~Riemann, {\em {Electroweak One Loop
  Corrections to the Decay of the Charged Vector Boson},}
  \href{https://dx.doi.org/10.1007/BF01441360}{Z.\  Phys.\  C {\bfseries 32}
  (1986) 121--125}.

\bibitem{Denner:1990tx}
A.~Denner and T.~Sack, {\em {The $W$-boson width},}
  \href{https://dx.doi.org/10.1007/BF01560267}{Z.\  Phys.\  C {\bfseries 46}
  (1990) 653--663}.

\bibitem{Dubovyk:2018rlg}
I.~Dubovyk, A.~Freitas, J.~Gluza, T.~Riemann, and J.~Usovitsch, {\em {Complete
  electroweak two-loop corrections to Z boson production and decay},}
  \href{https://dx.doi.org/10.1016/j.physletb.2018.06.037}{Phys.\  Lett.\  B
  {\bfseries 783} (2018) 86--94} {\ttfamily
  [\href{https://arxiv.org/abs/1804.10236}{arXiv:1804.10236}]}.

\bibitem{deBlas:2021wap}
J.~de~Blas, {\em et al.}, {\em {Global analysis of electroweak data in the
  Standard Model}.} {\ttfamily
  \href{https://arxiv.org/abs/2112.07274}{arXiv:2112.07274}}.

\bibitem{ParticleDataGroup:2020ssz}
{\bfseries Particle Data Group} Collaboration, {\em {Review of Particle
  Physics},} \href{https://dx.doi.org/10.1093/ptep/ptaa104}{PTEP {\bfseries
  2020} (2020) 083C01 and 2021 update}.

\bibitem{ATLAS:2020xea}
{\bfseries ATLAS} Collaboration, {\em {Test of the universality of $\tau$ and
  $\mu$ lepton couplings in $W$-boson decays with the ATLAS detector},}
  \href{https://dx.doi.org/10.1038/s41567-021-01236-w}{Nature Phys.\
  {\bfseries 17} (2021) 813--818} {\ttfamily
  [\href{https://arxiv.org/abs/2007.14040}{arXiv:2007.14040}]}.

\bibitem{Bernreuther:2016ccf}
W.~Bernreuther, L.~Chen, O.~Dekkers, T.~Gehrmann, and D.~Heisler, {\em {The
  forward-backward asymmetry for massive bottom quarks at the $Z$ peak at
  next-to-next-to-leading order QCD},}
  \href{https://dx.doi.org/10.1007/JHEP01(2017)053}{JHEP {\bfseries 01} (2017)
  053} {\ttfamily [\href{https://arxiv.org/abs/1611.07942}{arXiv:1611.07942}]}.

\bibitem{Ando:2011}
T.~Ando, {\em {Predictive bayesian model selection},}
  \href{https://dx.doi.org/10.1080/01966324.2011.10737798}{Am.\  J.\  Math.\
  Manag.\  Sci.\  {\bfseries 31} (2011) 13}.

\bibitem{deBlas:2017xtg}
J.~de~Blas, J.~C.~Criado, M.~Perez-Victoria, and J.~Santiago, {\em {Effective
  description of general extensions of the Standard Model: the complete
  tree-level dictionary},}
  \href{https://dx.doi.org/10.1007/JHEP03(2018)109}{JHEP {\bfseries 03} (2018)
  109}
{\ttfamily [\href{https://arxiv.org/abs/1711.10391}{arXiv:1711.10391}]}.

\bibitem{Minkowski:1977sc}
P.~Minkowski, {\em {$\mu \to e\gamma$ at a Rate of One Out of $10^{9}$ Muon
  Decays?}} \href{https://dx.doi.org/10.1016/0370-2693(77)90435-X}{Phys.\ \
  Lett.\ \ B {\bfseries 67} (1977) 421--428}.

\bibitem{GellMann:1980vs}
M.~Gell-Mann, P.~Ramond, and R.~Slansky, {\em {Complex Spinors and Unified
  Theories},} Conf.\ \ Proc.\ \ C {\bfseries 790927} (1979) 315--321 {\ttfamily
  [\href{https://arxiv.org/abs/1306.4669}{arXiv:1306.4669}]}.

\bibitem{Yanagida:1979as}
T.~Yanagida, {\em {Horizontal gauge symmetry and masses of neutrinos},} Conf.\
  \ Proc.\ \ C {\bfseries 7902131} (1979) 95--99.

\bibitem{Mohapatra:1979ia}
R.~N.~Mohapatra and G.~Senjanovic, {\em {Neutrino Mass and Spontaneous Parity
  Nonconservation},}
  \href{https://dx.doi.org/10.1103/PhysRevLett.44.912}{Phys.\ \ Rev.\ \ Lett.\
  {\bfseries 44} (1980) 912}.

\bibitem{Foot:1988aq}
R.~Foot, H.~Lew, X.~G.~He, and G.~C.~Joshi, {\em {Seesaw Neutrino Masses
  Induced by a Triplet of Leptons},}
\href{https://dx.doi.org/10.1007/BF01415558}{Z.\  Phys.\  {\bfseries C44}
  (1989) 441}.

\bibitem{Davies:1990sc}
A.~J.~Davies and X.-G.~He, {\em {Tree Level Scalar Fermion Interactions
  Consistent With the Symmetries of the Standard Model},}
  \href{https://dx.doi.org/10.1103/PhysRevD.43.225}{Phys.\  Rev.\  D {\bfseries
  43} (1991) 225--235}.

\bibitem{delAguila:2008pw}
F.~del Aguila, J.~de~Blas, and M.~Perez-Victoria, {\em {Effects of new leptons
  in Electroweak Precision Data},}
  \href{https://dx.doi.org/10.1103/PhysRevD.78.013010}{Phys.\  Rev.\
  {\bfseries D78} (2008) 013010}
{\ttfamily [\href{https://arxiv.org/abs/0803.4008}{arXiv:0803.4008}]}.

\bibitem{Willmann:1998gd}
L.~Willmann {\em et~al.}, {\em {New bounds from searching for muonium to
  anti-muonium conversion},}
  \href{https://dx.doi.org/10.1103/PhysRevLett.82.49}{Phys.\  Rev.\  Lett.\
  {\bfseries 82} (1999) 49--52} {\ttfamily
  [\href{https://arxiv.org/abs/hep-ex/9807011}{hep-ex/9807011}]}.

\bibitem{ATLAS:2020fzp}
{\bfseries ATLAS} Collaboration, {\em {A search for the dimuon decay of the
  Standard Model Higgs boson with the ATLAS detector},}
  \href{https://dx.doi.org/10.1016/j.physletb.2020.135980}{Phys.\  Lett.\  B
  {\bfseries 812} (2021) 135980} {\ttfamily
  [\href{https://arxiv.org/abs/2007.07830}{arXiv:2007.07830}]}.

\bibitem{ATLAS:2019nkf}
{\bfseries ATLAS} Collaboration, {\em {Combined measurements of Higgs boson
  production and decay using up to $80$ fb$^{-1}$ of proton-proton collision
  data at $\sqrt{s}=$ 13 TeV collected with the ATLAS experiment},}
  \href{https://dx.doi.org/10.1103/PhysRevD.101.012002}{Phys.\  Rev.\  D
  {\bfseries 101} (2020) 012002} {\ttfamily
  [\href{https://arxiv.org/abs/1909.02845}{arXiv:1909.02845}]}.

\bibitem{CMS:2020xwi}
{\bfseries CMS} Collaboration, {\em {Evidence for Higgs boson decay to a pair
  of muons},} \href{https://dx.doi.org/10.1007/JHEP01(2021)148}{JHEP {\bfseries
  01} (2021) 148} {\ttfamily
  [\href{https://arxiv.org/abs/2009.04363}{arXiv:2009.04363}]}.

\bibitem{CMS:2018uag}
{\bfseries CMS} Collaboration, {\em {Combined measurements of Higgs boson
  couplings in proton\textendash{}proton collisions at $\sqrt{s}=13\,\text
  {Te}\text {V} $},}
  \href{https://dx.doi.org/10.1140/epjc/s10052-019-6909-y}{Eur.\  Phys.\  J.\
  C {\bfseries 79} (2019) 421} {\ttfamily
  [\href{https://arxiv.org/abs/1809.10733}{arXiv:1809.10733}]}.

\bibitem{deBlas:2019rxi}
J.~de~Blas {\em et~al.}, {\em {Higgs Boson Studies at Future Particle
  Colliders},} \href{https://dx.doi.org/10.1007/JHEP01(2020)139}{JHEP
  {\bfseries 01} (2020) 139} {\ttfamily
  [\href{https://arxiv.org/abs/1905.03764}{arXiv:1905.03764}]}.

\bibitem{ATLAS:2021wob}
{\bfseries ATLAS} Collaboration, {\em {Search for new phenomena in three- or
  four-lepton events in $pp$ collisions at $\sqrt s$ =13 TeV with the ATLAS
  detector},} \href{https://dx.doi.org/10.1016/j.physletb.2021.136832}{Phys.\
  Lett.\  B {\bfseries 824} (2022) 136832} {\ttfamily
  [\href{https://arxiv.org/abs/2107.00404}{arXiv:2107.00404}]}.

\bibitem{ATLAS:2022yhd}
{\bfseries ATLAS} Collaboration, {\em {Search for type-III seesaw heavy leptons
  in leptonic final states in $pp$ collisions at $\sqrt{s} = 13$ TeV with the
  ATLAS detector}.} {\ttfamily
  \href{https://arxiv.org/abs/2202.02039}{arXiv:2202.02039}}.

\bibitem{CMS:2022nty}
{\bfseries CMS} Collaboration, {\em {Inclusive nonresonant multilepton probes
  of new phenomena at $\sqrt s$=13\,\,TeV},}
  \href{https://dx.doi.org/10.1103/PhysRevD.105.112007}{Phys.\  Rev.\  D
  {\bfseries 105} (2022) 112007} {\ttfamily
  [\href{https://arxiv.org/abs/2202.08676}{arXiv:2202.08676}]}.

\bibitem{Bhattiprolu:2019vdu}
P.~N.~Bhattiprolu and S.~P.~Martin, {\em {Prospects for vectorlike leptons at
  future proton-proton colliders},}
  \href{https://dx.doi.org/10.1103/PhysRevD.100.015033}{Phys.\  Rev.\
  {\bfseries D100} (2019) 015033}
{\ttfamily [\href{https://arxiv.org/abs/1905.00498}{arXiv:1905.00498}]}.

\bibitem{Muong-2:2002wip}
{\bfseries Muon g-2} Collaboration, {\em {Measurement of the positive muon
  anomalous magnetic moment to 0.7 ppm},}
  \href{https://dx.doi.org/10.1103/PhysRevLett.89.101804}{Phys.\  Rev.\  Lett.\
   {\bfseries 89} (2002) 101804} {\ttfamily
  [\href{https://arxiv.org/abs/hep-ex/0208001}{hep-ex/0208001}]}. [Erratum:
  Phys.Rev.Lett. 89, 129903 (2002)].

\bibitem{Muong-2:2004fok}
{\bfseries Muon g-2} Collaboration, {\em {Measurement of the negative muon
  anomalous magnetic moment to 0.7 ppm},}
  \href{https://dx.doi.org/10.1103/PhysRevLett.92.161802}{Phys.\  Rev.\  Lett.\
   {\bfseries 92} (2004) 161802} {\ttfamily
  [\href{https://arxiv.org/abs/hep-ex/0401008}{hep-ex/0401008}]}.

\bibitem{Muong-2:2006rrc}
{\bfseries Muon g-2} Collaboration, {\em {Final Report of the Muon E821
  Anomalous Magnetic Moment Measurement at BNL},}
  \href{https://dx.doi.org/10.1103/PhysRevD.73.072003}{Phys.\  Rev.\  D
  {\bfseries 73} (2006) 072003} {\ttfamily
  [\href{https://arxiv.org/abs/hep-ex/0602035}{hep-ex/0602035}]}.

\bibitem{Muong-2:2021ojo}
{\bfseries Muon g-2} Collaboration, {\em {Measurement of the Positive Muon
  Anomalous Magnetic Moment to 0.46 ppm},}
  \href{https://dx.doi.org/10.1103/PhysRevLett.126.141801}{Phys.\  Rev.\
  Lett.\  {\bfseries 126} (2021) 141801} {\ttfamily
  [\href{https://arxiv.org/abs/2104.03281}{arXiv:2104.03281}]}.

\bibitem{Aoyama:2020ynm}
T.~Aoyama {\em et~al.}, {\em {The anomalous magnetic moment of the muon in the
  Standard Model}.} {\ttfamily
  \href{https://arxiv.org/abs/2006.04822}{arXiv:2006.04822}}.

\bibitem{Czarnecki:2001pv}
A.~Czarnecki and W.~J.~Marciano, {\em {The Muon anomalous magnetic moment: A
  Harbinger for `new physics'},}
  \href{https://dx.doi.org/10.1103/PhysRevD.64.013014}{Phys.\ \ Rev.\ \ D
  {\bfseries 64} (2001) 013014} {\ttfamily
  [\href{https://arxiv.org/abs/hep-ph/0102122}{hep-ph/0102122}]}.

\bibitem{Kannike:2011ng}
K.~Kannike, M.~Raidal, D.~M.~Straub, and A.~Strumia, {\em {Anthropic solution
  to the magnetic muon anomaly: the charged see-saw},}
  \href{https://dx.doi.org/10.1007/JHEP02(2012)106}{JHEP {\bfseries 02} (2012)
  106} {\ttfamily [\href{https://arxiv.org/abs/1111.2551}{arXiv:1111.2551}]}.
  [Erratum: JHEP 10, 136 (2012)].

\bibitem{Dermisek:2013gta}
R.~Dermisek and A.~Raval, {\em {Explanation of the Muon g-2 Anomaly with
  Vectorlike Leptons and its Implications for Higgs Decays},}
  \href{https://dx.doi.org/10.1103/PhysRevD.88.013017}{Phys.\  Rev.\
  {\bfseries D88} (2013) 013017}
{\ttfamily [\href{https://arxiv.org/abs/1305.3522}{arXiv:1305.3522}]}.

\bibitem{Freitas:2014pua}
A.~Freitas, J.~Lykken, S.~Kell, and S.~Westhoff, {\em {Testing the Muon g-2
  Anomaly at the LHC},} \href{https://dx.doi.org/10.1007/JHEP09(2014)155,
  10.1007/JHEP05(2014)145}{JHEP {\bfseries 05} (2014) 145} {\ttfamily
  [\href{https://arxiv.org/abs/1402.7065}{arXiv:1402.7065}]}.
[Erratum: JHEP09,155(2014)].

\bibitem{Megias:2017dzd}
E.~Megias, M.~Quiros, and L.~Salas, {\em {$g_\mu-2$ from Vector-Like Leptons in
  Warped Space},} \href{https://dx.doi.org/10.1007/JHEP05(2017)016}{JHEP
  {\bfseries 05} (2017) 016} {\ttfamily
  [\href{https://arxiv.org/abs/1701.05072}{arXiv:1701.05072}]}.

\bibitem{Poh:2017tfo}
Z.~Poh and S.~Raby, {\em {Vectorlike leptons: Muon g-2 anomaly, lepton flavor
  violation, Higgs boson decays, and lepton nonuniversality},}
  \href{https://dx.doi.org/10.1103/PhysRevD.96.015032}{Phys.\ \ Rev.\ \ D
  {\bfseries 96} (2017) 015032} {\ttfamily
  [\href{https://arxiv.org/abs/1705.07007}{arXiv:1705.07007}]}.

\bibitem{Kowalska:2017iqv}
K.~Kowalska and E.~M.~Sessolo, {\em {Expectations for the muon g-2 in
  simplified models with dark matter},}
  \href{https://dx.doi.org/10.1007/JHEP09(2017)112}{JHEP {\bfseries 09} (2017)
  112} {\ttfamily [\href{https://arxiv.org/abs/1707.00753}{arXiv:1707.00753}]}.

\bibitem{Raby:2017igl}
S.~Raby and A.~Trautner, {\em {Vectorlike chiral fourth family to explain muon
  anomalies},} \href{https://dx.doi.org/10.1103/PhysRevD.97.095006}{Phys.\ \
  Rev.\ \ D {\bfseries 97} (2018) 095006} {\ttfamily
  [\href{https://arxiv.org/abs/1712.09360}{arXiv:1712.09360}]}.

\bibitem{Calibbi:2018rzv}
L.~Calibbi, R.~Ziegler, and J.~Zupan, {\em {Minimal models for dark matter and
  the muon g$-$2 anomaly},}
  \href{https://dx.doi.org/10.1007/JHEP07(2018)046}{JHEP {\bfseries 07} (2018)
  046} {\ttfamily [\href{https://arxiv.org/abs/1804.00009}{arXiv:1804.00009}]}.

\bibitem{Crivellin:2018qmi}
A.~Crivellin, M.~Hoferichter, and P.~Schmidt-Wellenburg, {\em {Combined
  explanations of $(g-2)_{\mu,e}$ and implications for a large muon EDM},}
  \href{https://dx.doi.org/10.1103/PhysRevD.98.113002}{Phys.\ \ Rev.\ \ D
  {\bfseries 98} (2018) 113002} {\ttfamily
  [\href{https://arxiv.org/abs/1807.11484}{arXiv:1807.11484}]}.

\bibitem{Kawamura:2019rth}
J.~Kawamura, S.~Raby, and A.~Trautner, {\em {Complete vectorlike fourth family
  and new U(1)' for muon anomalies},}
  \href{https://dx.doi.org/10.1103/PhysRevD.100.055030}{Phys.\ \ Rev.\ \ D
  {\bfseries 100} (2019) 055030} {\ttfamily
  [\href{https://arxiv.org/abs/1906.11297}{arXiv:1906.11297}]}.

\bibitem{Kawamura:2019hxp}
J.~Kawamura, S.~Raby, and A.~Trautner, {\em {Complete vectorlike fourth family
  with U(1)' : A global analysis},}
  \href{https://dx.doi.org/10.1103/PhysRevD.101.035026}{Phys.\ \ Rev.\ \ D
  {\bfseries 101} (2020) 035026} {\ttfamily
  [\href{https://arxiv.org/abs/1911.11075}{arXiv:1911.11075}]}.

\bibitem{Endo:2020tkb}
M.~Endo and S.~Mishima, {\em {Muon $g-2$ and CKM unitarity in extra lepton
  models},} \href{https://dx.doi.org/10.1007/JHEP08(2020)004}{JHEP {\bfseries
  08} (2020) 004} {\ttfamily
  [\href{https://arxiv.org/abs/2005.03933}{arXiv:2005.03933}]}.

\bibitem{Crivellin:2020ebi}
A.~Crivellin, F.~Kirk, C.~A.~Manzari, and M.~Montull, {\em {Global Electroweak
  Fit and Vector-Like Leptons in Light of the Cabibbo Angle Anomaly},}
  \href{https://dx.doi.org/10.1007/JHEP12(2020)166}{JHEP {\bfseries 12} (2020)
  166} {\ttfamily [\href{https://arxiv.org/abs/2008.01113}{arXiv:2008.01113}]}.

\bibitem{Athron:2021iuf}
P.~Athron, {\em et al.}, {\em {New physics explanations of a$_{\mu}$ in light
  of the FNAL muon g \ensuremath{-} 2 measurement},}
  \href{https://dx.doi.org/10.1007/JHEP09(2021)080}{JHEP {\bfseries 09} (2021)
  080} {\ttfamily [\href{https://arxiv.org/abs/2104.03691}{arXiv:2104.03691}]}.

\bibitem{Djouadi:2021wvb}
A.~Djouadi, {\em et al.}, {\em {New fermions in the light of the $(g-2)_\mu$}.}
  {\ttfamily \href{https://arxiv.org/abs/2112.12502}{arXiv:2112.12502}}.

\end{thebibliography}\endgroup
\end{document}